\documentclass[preprint,showpacs,titlepage,aps,prd,
tightenlines,amsmath,byrevtex,nofootinbib]{revtex4}

\usepackage{graphicx}

\def\lsim{\raise0.3ex\hbox{$\;<$\kern-0.75em\raise-1.1ex
\hbox{$\sim\;$}}}
\def\gsim{\raise0.3ex\hbox{$\;>$\kern-0.75em\raise-1.1ex
\hbox{$\sim\;$}}}

\begin{document}

\preprint{hep-ph/0612002}

\title{ On {\em in situ} Determination of Earth Matter Density in \\ Neutrino Factory}

\author{Hisakazu Minakata}
\email{minakata@phys.metro-u.ac.jp}
\author{Shoichi Uchinami}
\email{uchinami@phys.metro-u.ac.jp}
\affiliation{Department of Physics, Tokyo Metropolitan University \\
1-1 Minami-Osawa, Hachioji, Tokyo 192-0397, Japan}

%\date{\today}
\date{February 16, 2007}

\vglue 1.4cm

\begin{abstract}
We point out that an accurate {\em in situ} determination of the 
earth matter density $\rho$ is possible in neutrino factory 
by placing a detector at the magic baseline, 
$L = \sqrt{2} \pi / G_{F} N_{e}$ where $N_{e}$ denotes electron number density. 
The accuracy of matter density determination is excellent in a region of 
relatively large $\theta_{13}$ with fractional uncertainty $\delta \rho / \rho$ of 
about 0.43\%, 1.3\%, and $\lsim$3\%  at 1$\sigma$ CL at 
$\sin^2 2\theta_{13}=0.1$, $10^{-2}$, and $3 \times 10^{-3}$, respectively. 
At smaller $\theta_{13}$ the uncertainty depends upon the CP phase 
$\delta$, but it remains small, 3\%-7\% in more than 3/4 of the entire 
region of $\delta$ at $\sin^2 2\theta_{13} = 10^{-4}$. 
The results would allow us to solve the problem of obscured CP violation 
due to the uncertainty of earth matter density in a wide range of 
$\theta_{13}$ and $\delta$. 
It may provide a test for the geophysical model of the earth, or 
it may serve as a method for stringent test of the MSW theory of 
neutrino propagation in matter once an accurate geophysical estimation 
of the matter density is available. 

\end{abstract}


\maketitle

%%%%%%%%%%%%%%%%%%%%%%%%

\section{Introduction}

In the past 10 years, the atmospheric \cite{SKatm}, 
the solar \cite{solar}, and the reactor experiments \cite{KamLAND} 
discovered neutrino oscillation and/or flavor conversion 
and identified neutrino masses and the lepton flavor mixing 
as the cause of the phenomena 
\cite{atm_evidence,KL_evidence,K2K_evidence,MINOS}. 
The next goal of the neutrino oscillation experiments is to explore 
the yet unknown 1-3 sector of the flavor mixing matrix, 
the MNS matrix \cite{MNS}, namely $\theta_{13}$, the CP phase $\delta$, 
and the neutrino mass hierarchy. 
If $\theta_{13}$ is relatively large, $\sin^2 2\theta_{13} \gsim 0.01$, 
most probably conventional superbeam \cite{superbeam} can do the job. 
But, if it turns out to be small, $\sin^2 2\theta_{13} \lsim 0.01$, 
it is likely that we need a new technology, either the neutrino factory 
\cite{nufact} or the beta beam \cite{beta}.  
Both options are under active considerations 
\cite{nufact_eu-us,beta2,international}.

In measurement in neutrino factory, 
a long baseline distance of several thousand km to a detector 
is preferred because the event rate increases as muon 
beam energy gets higher  and one wants to keep $L/E$ not too small for 
oscillation signatures \cite{golden}. 
Then, according to the Mikheyev-Smirnov-Wolfenstein (MSW) theory 
of neutrino propagation in matter \cite{wolfenstein,MS}, 
the earth matter effect is one of the crucial ingredients in precision 
measurement of the lepton mixing parameters. 
It may acts as a serious background in determination of CP 
violating phase $\delta$ \cite{AKS,MNprd98,Freund,MNjhep01}. 
In fact, it has been debated to what extent uncertainty in 
the earth matter density affects accuracies of determination 
of lepton mixing parameters, in particular, the CP phase $\delta$. 
See, for example, 
\cite{Joe,Burguet-C,yasuda,munich1,Ohlsson,munich2}.

One way of dealing with the problem of uncertainty of the matter density 
is to trust geophysical estimation. 
It appears that there is a consensus in geophysics community that 
the uncertainty in the density in the mantle region of the earth is 
less than 5\% \cite{geller-hara}. 
If the uncertainty can be regarded as robust, then the uncertainty 
may not be so crucial to hurt the accuracy of measurement of 
the mixing parameters, though it still produces sizable uncertainties. 
Notice, however, that the authors of \cite{yasuda} take much more 
conservative attitude in estimating the uncertainty of the matter density. 
See e.g., \cite{munich2} for a recent estimation of impact of the 
matter density uncertainty on accuracy in mixing parameter determination.

However, it would be much nicer if neutrino factory experiments 
can measure {\em in situ} the average matter density in the mantle 
region of the earth. 
It is not completely satisfactory that the most advanced apparatus 
for precision measurement of the lepton mixing parameters 
has to rely on the parameter that cannot be directly measured. 
In fact, the same attitude is shared by the authors of the pioneering work of 
neutrino factory analysis \cite{golden}, in which the accuracy of 
matter density determination by neutrino factory is estimated.
Describing an improved way for the {\em in situ} measurement of the 
earth matter density in neutrino factory is nothing but the goal of this paper. 
The recent progress in understanding of how to lower the 
analysis energy threshold \cite{Cervera_nufact06} will be the key to the 
remarkable accuracy we will uncover.

Moreover, measured matter density in deep interior of the earth by 
neutrinos would offer an independent test for the theory of earth 
structure and its formation. 
Note that the mantle region contains 70\% of the earth mass. 
If it is significantly different from the value predicted by geophysical 
models 
(after taking account of the uncertainties in both prediction and measurement), 
we have to think about at least one of the following possibilities: 
(1) A drastic revision of geophysical model of the earth must be attempted. 
(2) The MSW theory of neutrino propagation is in error. 
We hope that this discussion is illuminative of the importance of the 
neutrino measurement of the earth matter density.

The chondrite earth model with constraint of seismic data results in 
earth model with several layers and the matter density in each layers 
can be predicted with reasonable precision with constraint of 
accurately measured earth mass. 
In particular, the density in the mantle region seems to be obtained 
with the least uncertainties \cite{Panasyuk}.  
If the geophysical estimation can be regarded as robust 
neutrino factory measurement of matter density we propose 
in this paper severely tests the theory of neutrino propagation 
in matter. 
Notice that the matter density at the solar core is currently measured 
by using neutrinos only up to a factor of $\sim$2 uncertainty \cite{bari_focus}.
Neutrino tomography of the earth by using accelerator and other neutrino 
sources has been discussed in many literatures \cite{tomography_beam,tomography_core,tomography_SN,tomography_rev}.

In Sec.~\ref{basic}, we give three conditions for accurate measurement 
of the earth matter density based on general considerations. 
In Sec.~\ref{scan}, we discuss what is the baseline that is most sensitive 
to matter density change.
In Sec.~\ref{binning}, we describe a concrete way of determining the 
earth matter density by neutrino factory. 
In Sec.~\ref{method}, we define the statistical method for analysis. 
In Sec.~\ref{results}, we describe the results of our analysis. 
In Sec.~\ref{delta-dep}, we address the problem of $\delta$-dependence 
of the error of matter density determination. 
In Sec.~\ref{iterative}, an iterative procedure for combined analysis 
of data at the intermediate and the far detectors is described. 
In Sec.~\ref{conclude}, we give concluding remarks.

\section{General considerations}
\label{basic}

Experiments which aim at measuring the earth matter density 
accurately by using neutrino oscillation must satisfy a few crucial requirements. 
Firstly, the matter effect must be sizable, at least comparable 
with the vacuum oscillation effect. 
The relative importance between the vacuum and the matter effects 
in long-baseline (LBL) neutrino oscillation experiments may be 
quantified by comparing the following two dimensionless quantities:
\begin{eqnarray}
\Delta_{31} &\equiv& \frac{|\Delta m^2_{31}| L}{4 E} = 1.27   
\left(\frac{|\Delta m^2_{31}|}{10^{-3}\mbox{eV}^2}\right)
\left(\frac{L}{1000 \mbox{km}}\right)
\left(\frac{E}{1\mbox{GeV}}\right)^{-1}
\\
aL &\equiv& \frac{1}{\sqrt{2}} G_F N_e L = 0.27 
\left(\frac{\rho}{2.8 \text{g/cm}^3}\right)
\left(\frac{L}{1000 \mbox{km}}\right) 
\label{vacuum-matter}
\end{eqnarray}
where $a$ is related to neutrino's index of refraction 
with $G_F$ being the Fermi constant and 
$N_e$ the electron number density in the earth. 
$N_e$ is related to the matter density $\rho$ as 
$N_e = Y_{e} N_{A} \rho$ with Avogadro's number $N_{A}$ and 
the electron fraction $Y_{e}$. We take $Y_{e}=0.5$. 
In most cases, the experimental setup is such that 
$\Delta_{31} \sim \pi/2$ to maximize the appearance signal. 
In view of (\ref{vacuum-matter}), 
the matter effect plays no significant role in settings 
of baseline less than $\sim 1000$ km.

Sizable matter effect is certainly the necessary but  {\it not} 
the sufficient condition for experiments to measure the matter density. 
What is important is the sensitivity to artificial matter density change 
$\delta \rho$. 
To quantify the sensitivity one has to have a variable experimental 
parameters through which one can control relative importance 
of the matter effect to that of the vacuum oscillation. 
Since $aL / \Delta_{31}$ is proportional to $E$  the natural candidate 
for relevant experimental parameters to vary would be the 
neutrino energy.

It is not the end of the list of requirements. 
A measurement at a given set up determines a combination of 
the mixing parameters, typically $\theta_{13}$, $\delta$, and the 
matter effect coefficient $a$.
Unless the former two parameters are separately measured in 
experiments in vacuum,\footnote{
%%%%%%%%%%%%%% footnote %%%%%%%%%%%%%%%%%
MEMPHYS \cite{MEMPHYS} would be an ideal apparatus for this 
purpose because even the T2K experiment \cite{T2K} is 
contaminated by the matter effect which leads to some sensitivities 
to the mass hierarchy in a limited region of $\delta$, 
as anticipated \cite{MNjhep01,nufact01_mina} and proved 
by the recent analyses \cite{T2KK}.  
}
the coefficient $a$ or the matter density 
alone cannot be cleanly determined by any single measurement. 
Generally speaking, the condition is hard to be met because 
most of the projects to determine unknowns in the lepton flavor mixing 
are designed to be sensitive not only to the matter effect, which is 
required to determine the mass hierarchy, but also to the 
CP phase $\delta$.

In summary, we list here the three requirements (A)$-$(C); 
(A) sensitivity to the matter effect. 
(B) existence of $\delta \rho$ sensitive tunable parameters. 
(C) mixing parameter independent measurement of the matter density. 
We will discuss  in the following sections below how these requirements  
can be satisfied.

\section{Which baseline?; preliminary consideration with energy scan}
\label{scan}

In this paper, we focus on neutrino factory measurement 
of the earth matter density. 
Since most part of the neutrino trajectory in neutrino factory is in 
the mantle region of the earth, what we mean by the matter density 
in this paper is that of the mantle region. 
Precisely speaking, the matter density we try to determine 
is an averaged value along the neutrino trajectory.

The first question we must address is; What is the appropriate 
baseline distance for the measurement? 
We try to find an answer to this question by choosing muon beam 
energy as the tunable parameter, the energy scan. 
Namely, we consider the neutrino factory measurement of numbers 
of appearance events $\nu_{e} \rightarrow \nu_{\mu}$ at 
two different muon beam energies. 
It will give us the normalized event number difference 
$\Delta N / N$. 
The question we address here is what is the baseline $L$ 
that gives rise to the strongest response in $\Delta N / N$ 
to matter density change.

For simplicity, we employ in this section the asymptotic expansion 
of the oscillation probability assuming $\Delta_{31} \ll 1$.\footnote{
%%%%%%%%%%%%%% \footnote %%%%%%%%%%%%%%%
In a standard setting, $\Delta_{31}$ is small; 
$\Delta_{31} = 0.24 
\left(\frac{\Delta m^2}{2.5 \times 10^{-3}\mbox{eV}^2}\right)
\left(\frac{L}{3000 \mbox{km}}\right)
\left(\frac{E}{\mbox{40 GeV}}\right)^{-1}
$
}
The standard formula for $\nu_{e} \rightarrow \nu_{\mu}$ appearance 
probability \cite{golden} 
(for its full expression, see (\ref{emuP}) in Sec.~\ref{delta-dep}) 
takes the form at high energies, 
$\Delta_{31} \equiv \left(\frac{\Delta m^2_{31} L}{4 E}\right) \ll 1$, as 
\begin{eqnarray}
P(\nu_{e} \rightarrow \nu_{\mu}) = 
\frac{A}{E^2} + \frac{B}{E^3}
\label{Pmue}
\end{eqnarray}
where $E$ is the neutrino energy and $L$ is the baseline distance. 
The coefficients $A$ and $B$ are given to leading order in $s_{13}$ by 
\begin{eqnarray}
A &=& 
\left( \frac{ \sin aL }{ aL } \right)^2 
\left( \frac{|\Delta m^2_{31}| L}{4}  \right)^2 
\left[
4 s^2_{23} s^2_{13} \pm 
4 \sin 2\theta_{12} c_{23} s_{23} \epsilon s_{13} \cos \delta + 
c_{23}^2 \sin^2 2\theta_{12} \epsilon^2 
\right] 
\nonumber \\
B &=& \pm 4 
\left( \frac{ \sin aL }{ aL } \right)^2 
\left( \frac{|\Delta m^2_{31}| L}{4}  \right)^3 
g(aL) 
\left[
2 s^2_{23} s^2_{13} +  
 \sin 2\theta_{12} c_{23} s_{23} \epsilon s_{13} \cos \delta 
 \left( 1 \pm \frac{\tan \delta}{g(aL)}  \right)  
\right],  
\label{coefficient}
\end{eqnarray}
where $\epsilon \equiv \vert \frac {\Delta m^2_{21}}  {\Delta m^2_{31}} \vert$, 
and $a = G_F N_e /\sqrt{2}$ as before. 
The function $g$ is defined by 
$g(x) =  1/x - \cot x$. 
The positive and the negative sign in (\ref{coefficient}) are for the 
normal and the inverted mass hierarchies, respectively. 
In (\ref{coefficient}) and throughout this paper, we use the 
constant matter density approximation.

It is well known that the integrated electron neutrino flux 
$F_{\nu e}$ from muon decay has energy distribution as 
$F_{\nu e} (E) = F_{0} y^{2} (1-y)$ 
(in units of number of $\nu$'s per unit area)  
where $y \equiv \frac{E}{E_{\mu}}$ and 
$F_{0} = \frac{12}{\pi} \frac{n_{\mu} E_{\mu}}{ L^2 m_{\mu}^2}$
%%%%%%%%%%%%%%% flux %%%%%%%%%%%%%%%%
with $n_{\mu}$ and $L$ being the number of useful muon decay 
and the baseline distance, respectively. 
Then, the number of wrong sign muon events is given, using 
the number of target atom $N_{T}$, 
and by approximating the energy dependence of the charged 
current (CC) cross section as linear, 
$\sigma_{CC} = \sigma_{0} E$, as 
\begin{eqnarray}
N(E_{\mu}) = N_{T} \int^{E_{\mu}}_{E_{th}} dE 
F_{e} (E) \sigma_{CC} (E) 
P(\nu_{e} \rightarrow \nu_{\mu}; E) = 
\frac{2}{\pi} \frac{n_{\mu} }{ L^2 m_{\mu}^2}
\sigma_{0} N_{T}
\left( E_{\mu} A +  3 B \right) 
\label{Nevent}
\end{eqnarray}
where we have assumed that the muon's threshold energy $E_{th}$ 
can be made low enough to allow the approximation 
$1 \gg \frac{E_{th}}{E_{\mu}} \simeq 0$.

Let us consider measurement at two adjacent muon beam energies 
$E_{\mu} $ and $E_{\mu} + \Delta E_{\mu}$ with 
number of events $N(E_{\mu})$ and $N(E_{\mu} + \Delta E_{\mu})$, 
and define their difference as 
$\Delta N \equiv N(E_{\mu} + \Delta E_{\mu}) - N(E_{\mu}) \simeq 
\frac {d N(E_{\mu})}{d E_{\mu}} \Delta E_{\mu}$. 
We define the double ratio 
$\frac{\Delta N}{N} / \frac{\Delta E_{\mu}}{E_{\mu}} $, 
which is independent of the uncertainties in the CC cross section, 
the neutrino flex, the baseline, and the number of target atoms. 
We obtain 
\begin{eqnarray}
\frac {\frac{\Delta N}{N}} {\frac{\Delta E_{\mu}}{E_{\mu}}} = 
\frac{1}{1 + 3 \frac{B}{E_{\mu} A} }.
\label{Dratio}
\end{eqnarray}

Now, the crucial question we have to ask is: What is the baseline 
distance for which variation of the matter density, or $a$, induces 
maximal changes in the double ratio? 
From (\ref{Dratio}), it is the place where $B/A$ changes significantly 
when $aL$ is varied. 
The answer is $aL = \pi$ at which  
$g(aL)$ (which is proportional to $B/A$) diverges. 
This distance is nothing but the one called the ``magic baseline'' 
in the literature \cite{magicBL}.
It is interesting to see that the magic baseline appears in our treatment 
as a point where the sensitivity to matter density variation is maximal. 
The length has been known in the theory of neutrino propagation 
in matter as ``refraction length'' \cite{wolfenstein}. 
For a recent discussion on the meaning of the magic baseline, 
see \cite{smirnov_magic}.

Despite that the asymptotic expansion is not valid at around 
the distance where $g(aL)$ diverges, one can show by using the 
full expression of the oscillation probability that the distance 
comparable to the magic baseline is indeed the most sensitive 
place for the double ratio to change in matter density. 
(See Fig.~5.1 of \cite{uchinami}.)  
Thus, the asymptotic expansion is a correct indicator for the right 
baseline distance for measurement of the matter density. 
%%%%%%%%%%%%%% ? footnote ? %%%%%%%%%%%%%%%%
(It is reminiscent of the feature that the one-loop QCD coupling 
constant diverges at the hadronic scale $\Lambda$.) 

\section{Measuring earth matter density at the magic baseline; Energy binning}
\label{binning}

We now switch to a different strategy of using neutrino energy 
as the tunable parameter, though we will make a brief comment 
on the method of energy scan at the end of Sec.~\ref{results}. 
We consider measurement of the earth matter density at the magic baseline 
\begin{eqnarray}
L = \frac{\sqrt{2} \pi} {G_F N_e}  = 7480 
\left(\frac{\rho}{4.2 \text{g/cm}^3}\right)^{-1} \text{km}. 
\label{magicL}
\end{eqnarray}
As was emphasized in the original article \cite{magicBL}, 
one of the most characteristic feature of the magic baseline is that 
the oscillation probability $P(\nu_{e} \rightarrow \nu_{\mu}) $ is 
independent of the CP violating phase $\delta$ \cite{BMW}. 
Then, one can measure $\theta_{13}$ independently of $\delta$, and 
this property has been utilized to resolve the parameter degeneracy 
by combining with detector at $L=3000-4000$ km 
in neutrino factory measurement of $\theta_{13}$ and $\delta$ 
\cite{magicBL,munich2}. 
We will show below that the property is also the key to our method 
for accurate measurement of the earth matter density, fulfilling 
the requirement (C) at least partly. 
We, however, encounter below the problem of unexpected 
$\delta$-dependence in a limited region of $\theta_{13}$ and 
$\delta$. See Sec.~\ref{delta-dep}.

Now, the key question is: 
What is the most efficient way to measure the matter density 
accurately at the magic baseline?
To formulate the right strategy for this purpose we analyze 
the structure of the appearance probability 
$P( \nu_e \rightarrow \nu_{\mu} )$. 
At around the magic baseline the Cervera {\it et al.} formula \cite{golden} 
has a very simple form with vanishingly small solar and interference terms 
(the first term in (\ref{emuP}) in Sec.~\ref{delta-dep}), 
\begin{eqnarray}
P( \nu_e \rightarrow \nu_{\mu} ) = 
s^2_{23} \sin^2 2\theta_{13}
\left[ \frac{\Delta_{31}\sin({aL \mp \Delta_{31}})}{(aL \mp \Delta_{31})} \right]^2 
\label{Pemu}
\end{eqnarray}
with $aL = \pi$, where the $\pm$ sign in (\ref{Pemu}) corresponds 
to the neutrino and the anti-neutrino channels. 
We note that when the matter density is perturbed, 
$\rho \rightarrow \rho + \delta \rho$, or 
$aL = \pi + \epsilon$ at the magic baseline, 
the response of the function in the square bracket in (\ref{Pemu}) 
is given as follows: 
\begin{eqnarray}
\frac{\sin({aL \mp \Delta_{31}})}{(aL \mp \Delta_{31})} = 
\frac{\sin({\pi \mp \Delta_{31}})}{(\pi \mp \Delta_{31})} 
\left[
1 - \epsilon g(\pi \mp \Delta_{31}) 
\right]
\label{response}
\end{eqnarray}
where $g(x) \equiv 1/x - \cot x$ as before.\footnote{
%%%%%%%%%%%%%%%% footnote %%%%%%%%%%%%%%%%%%
It may be instructive to remark that $g(x)$ is a monotonically increasing 
function of $x$ in a range $2n\pi < x < (2n+1)\pi$ where $n$ is an integer. 
It has zero at $x = 0$ and at $x = 4.493$, and diverges to $\pm \infty$ at 
$x=n\pi \mp \epsilon$. 
}
Equation (\ref{response}) indicates that the response to density 
change depends upon the neutrino energy, 
the channel ($\nu$ or $\bar{\nu}$), and the neutrino mass ordering, 
the normal ($\Delta_{31} > 0$), or the inverted ($\Delta_{31} < 0$) hierarchies.

For definiteness, we first discuss the case of the normal hierarchy 
($\Delta_{31} > 0$). 
In the neutrino channel, the equation $g(\pi - \Delta_{31})=0$ has a 
solution at $\Delta_{31} - \pi=0$. 
Namely, the critical energy is given by $E_{c} \simeq 7.6$ GeV for 
$L=7500$ km. 
Since $g(x)$ is an odd function of $x$, the probability decreases (increases) 
at neutrino energies  $E > E_{c}$ ($E < E_{c}$) as the density increases 
in the neutrino channel. 
In the anti-neutrino channel, the equation $g(\pi + \Delta_{31})=0$ has a 
solution at $\Delta_{31} + \pi = 4.493$, and $g(x) > 0$ for $x > 4.493$ 
and vice versa. 
The corresponding critical energy is $E_{c} \simeq 17.6$ GeV. 
The antineutrino probability increases (decreases) 
at neutrino energies  $E > E_{c}$ ($E < E_{c}$) as the density increases, 
a behavior opposite  to the neutrino channel. 
In the inverted hierarchy, the situation of the neutrino and the anti-neutrino 
channels completely reverses, as one can easily confirm.

%%%%%%%%%%%% FIGURE I %%%%%%%%%%%%%%%
\begin{figure}[bhtp]
\vglue 0.2cm
\begin{center}
%\hglue  -0.2cm
\includegraphics[width=1.0\textwidth]{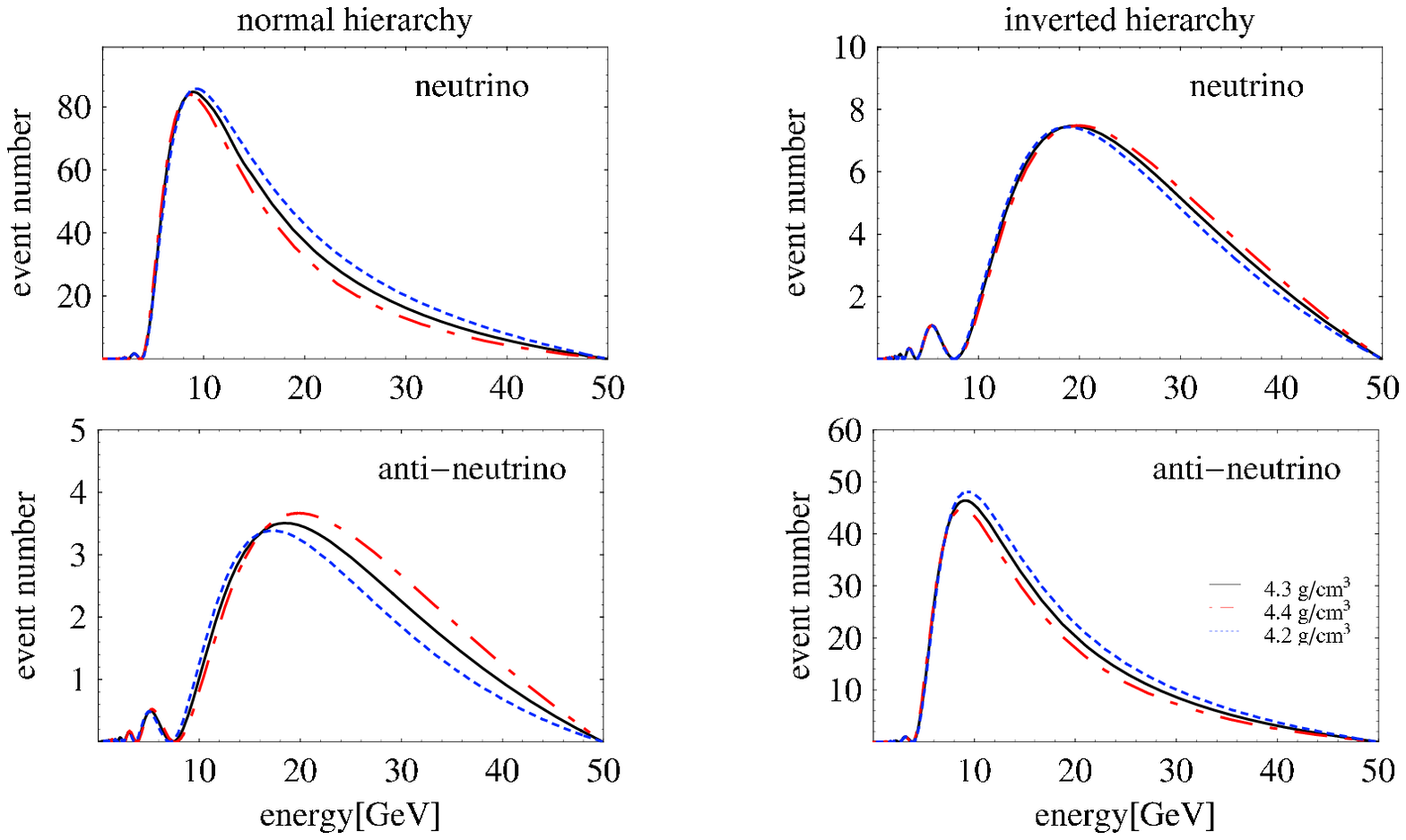}
\end{center}
%\vglue -0.9cm
\caption{
The energy distribution of event number (per GeV) is plotted with 
three values of the matter density, 
4.2 g/cm$^3$ (shown in blue dotted curve), 
4.3 g/cm$^3$ (black solid), and 
4.4 g/cm$^3$ (red dash-dotted). 
The left and the right two panels in Fig.~\ref{spectrum} are for 
the cases of the normal and the inverted mass hierarchies, respectively. 
The mixing parameters are taken as $\delta =0$ and 
$\sin^2 2 \theta_{13} = 0.01$. 
}
\label{spectrum}
\end{figure}
%%%%%%%%%%%% FIGURE I %%%%%%%%%%%%%%%

In Fig.~\ref{spectrum}, the event number distributions are plotted as 
a function of neutrino energy with three values of the matter densities. 
4.2, 4.3, and 4.4 g/cm$^3$. 
The mixing parameters are taken as $\delta =0$ and 
$\sin^2 2 \theta_{13} = 0.01$. 
The left and the right two panels of Figs.~\ref{spectrum} are for the cases 
of the normal and the inverted mass hierarchies, respectively. 
Notice that the oscillation probability is the only piece that is 
sensitive to matter density variation, and hence our above estimation 
of the critical energy should apply to the event number distribution as well. 
Figure~\ref{spectrum} confirms the qualitative behavior expected 
by the above analysis of response of the oscillation probability 
to the matter density perturbation. 
The difference between the critical energies in the neutrino and the 
anti-neutrino channel, as well as the characteristic feature is obvious; 
Smaller (larger) number of events in the neutrino (anti-neutrino) 
channel at energies above $E_{c}$ for higher matter density in the normal 
hierarchy case 
(and a completely reversed feature in the inverted hierarchy case). 
Therefore, it appears that one can expect the largest sensitivity to 
the matter density by using the two energy bins below and above $E_{c}$. 
The opposite response of the neutrino and the antineutrino channels to 
the matter density variation suggests that higher sensitivity can be 
achieved by using both channels, which will be verified in 
Sec.~\ref{results}.

We emphasize that lowering the analysis threshold, 
the possibility under active investigation \cite{Cervera_nufact06}, 
is the key to the feasibility of our method of energy binning, 
because the information below $E \lsim 10$ GeV is quite 
essential for neutrino (anti-neutrino) channel for the normal 
(inverted) hierarchy. 
We used the formula (\ref{Nevent}) but with the exact expression 
of the oscillation probability \cite{KTY} to compute the number of events 
with suitable modification of the range of integration over neutrino energy.  
In our analysis in this paper we use only the signal events by 
the charged current (CC) reaction. 
We take the cross section of CC reactions (including deep inelastic reactions) 
$\sigma_{CC} = 0.67 \times 10^{-38} (E / \text{GeV})$ cm$^2$ 
for neutrinos and 
$\sigma_{CC} = 0.34 \times 10^{-38} (E / \text{GeV})$ cm$^2$ 
for anti-neutrinos \cite{golden} in our analysis. 
%%%%%%%%%% MIXING PARAMETER %%%%%%%%%%%%%
Throughout this paper, apart from the comment at the end of 
Sec.~\ref{fixed-delta}, the true values of the neutrino 
mixing parameters are assumed as follows: 
$|\Delta m^2_{31}| = 2.5 \times 10^{-3} \text{eV}^2$, 
$\Delta m^2_{21} = 7.9 \times 10^{-5} \text{eV}^2$, 
$\sin^2 2 \theta_{12} = 0.86$, and  
$\sin^2 2 \theta_{23} = 1$.

In principle the disappearance channel $\nu_{\mu} \rightarrow \nu_{\mu}$ 
can be added to the analysis within the setting of magnetized iron detector. 
We do not go into this task in this paper; 
Though the matter effect is known to be sizable at these distances 
\cite{Gandhi} 
the response to matter density change does not appear to be favorable.

\section{Analysis method}
\label{method}

\subsection{Experimental setting}
\label{setting}

In this paper, we consider a setting of neutrino factory with muon energy 
$E_{\mu}=50$ GeV and assume total $3 \times 10^{21}$ useful 
muon decays per each polarity, as quoted in \cite{Blondel}. 
It may require 5-10 years operation of muon storage ring 
depending upon its luminosity.\footnote{
%%%%%%%%%%%%%%%% footnote %%%%%%%%%%%%%%%%%
The authors of \cite{munich1,munich2} assume more modest value of 
total $2 \times 10^{21}$ useful muon decays per each polarity. 
On the other hand, if the ``reference neutrino factory'' setting \cite{Long_nufact06} 
is realized one may be able to accumulate 
$5 \times 10^{21}$ muon decays by 5 years operation per each polarity. 
}
Our basic attitude is that operation of the neutrino factory will 
be coordinated by the requirement of optimization for 
accurate measurement of CP phase $\delta$ and $\theta_{13}$. 
Therefore. measurement of the matter density should be carried out 
as a ``byproduct'', within a given schedule optimized for the above purpose.

A standard setting of the neutrino factory includes two magnetized 
iron detectors, one at baseline $L=3000-4000$ km and the other 
at the magic baseline $L=7500$ km, which will be denoted as the 
intermediate and the far detectors, respectively. 
Note that the distance to the far detector does not exactly coincide 
with the magic baseline as given in (\ref{magicL}) 
if the matter density is $4.3 \text{g/cm}^2$, for example. 
But, the discrepancy of this level should be admitted because 
it is unlikely that we can find a site for the far detector at the 
distance equal to the magic baseline in a mathematical precision. 
Note also that for $L=7500$ km the maximum depth of the neutrino 
trajectory is about 1220 km  in the lower mantle region of the earth. 
Hence, neutrino passes through the lower mantle region in more 
than half a fraction of its trajectory. Note also that for $L=7500$ km 
neutrino passes through the lower mantle region in more 
than half a fraction of its trajectory and its maximum depth is about 1220 km.

In this paper, we analyze only the data taken by the far detector, 
leaving a full analysis of the data at the intermediate and the far 
detectors for future work. 
As a way of effectively implementing the information obtained by 
the intermediate detector, we impose the constraint on $\delta$ in 
doing fit in the analysis of far detector data. 
The fiducial mass of the far detector is assumed to be 40 kton. 
We also assume that the efficiency of event reconstruction is 
80\% and is independent of energy, which appears to be a good 
approximation to the efficiency in most of the relevant energy 
region shown in \cite{Cervera_nufact06}.\footnote{
%%%%%%%%%%%%%%%% footnote %%%%%%%%%%%%%%%%%%
The efficiency presented in these references is a preliminary one 
and is being under further improvement. 
The currently estimated efficiency may be lower than 80\% at 
energies at around 5 GeV. However, our optimistic assumption 
may not hurt the sensitivity estimate significantly because 
the event numbers which comes from this region is rather small. 
}
%%%%%%%%%%%%%%%% footnote %%%%%%%%%%%%%%%%%%
See also \cite{Ellis_nufact06} for the plot.

We now argue that analyzing only the data of the far detector is a 
sensible first approximation. 
To a good approximation there is a separation between the 
intermediate and the far detectors about their functions; 
The far detector has no sensitivity to the CP violating phase $\delta$, 
and the sensitivity to it solely comes from the intermediate detector. 
The matter effect which is crucial to determine the earth matter density 
is mainly felt by the far detector. 
Though $\theta_{13}$ is determined jointly by the intermediate and 
the far detectors, the latter is crucial to resolve degeneracy to 
precisely measure $\theta_{13}$ and $\delta$ \cite{magicBL}. 
%(See Secs.~\ref{results} for more about this point.) 
%
In fact, we will show that the maximally achievable accuracy of 
determination of the matter density only with the far detector is 
already quite remarkable, 
$\simeq$1\% ($\simeq$2\%) level for $\sin^2 2\theta_{13}=0.01$ (0.001) 
at 1 $\sigma$ CL, as will be shown in Sec.~\ref{results}. 
%The sensitivity in the case of the inverted hierarchy, however, is somewhat reduced. 

\subsection{Statistical procedure for quantitative analysis}
\label{procedure}

In our analysis, we use two energy bins, 
$5~\text{GeV} \leq E \leq 10~\text{GeV}$ and 
$10~\text{GeV} \leq E \leq 50~\text{GeV}$ for neutrino,  and 
$5~\text{GeV} \leq E \leq 20~\text{GeV}$ and 
$20~\text{GeV} \leq E \leq 50~\text{GeV}$ for anti-neutrino 
channels for the reasons discussed in Sec.~\ref{binning}. 
Notice that the boundary between the low- and the high-energy 
bins are somewhat different between the neutrino and the 
anti-neutrino channels, as can be seen to be appropriate 
in Fig.~\ref{spectrum}. 
While we do not explicitly deal with the issue of finite resolution in 
the reconstructed neutrino energy it will be partially taken care of 
by the systematic error discussed below.

The definition of $\Delta\chi^2$ for our analysis is given by 
\begin{eqnarray}
 \Delta\chi^2
 &\equiv&
 \min_{\text{$\alpha$'s}}
  \sum_{a = \nu, \bar{\nu}}
          \left[
           \sum_{i=1}^{2}
            \left\{
	     \frac{
	      \left( N_{ai}^{obs}
	             - ( 1 + \alpha_i + \alpha_a + \alpha ) N^{exp}_{ai}
	      \right)^2 }
	      { N^{exp}_{ai}
                + \sigma_{pb}^2 (N^{exp}_{ai})^2 }
             + \frac{ \alpha_i^2 }{ \sigma_{Pb}^2 }
	    \right\}
	   + \frac{ \alpha_a^2 }{ \sigma_{pB}^2 }
	  \right]
	  + \frac{ \alpha^2 }{ \sigma_{PB}^2 },
\label{chi2}
\end{eqnarray}
where $N_{ai}^{obs}$ is the number of events in $i$-th bin 
computed with the values of input parameters, 
and $N_{ai}^{exp}$ is the one computed with 
certain trial set of parameters. 
The form of $\Delta\chi^2$ with the notation of errors is analogous to 
that used in \cite{reactorCP}. 
The generalization to measurement at multiple energy bins, 
if necessary, is straightforward. 

The nature of the errors $\sigma_{PB}$, $\sigma_{Pb}$, $\sigma_{pB}$, 
and $\sigma_{pb}$ and the examples of them are as follows. 
For simplicity of notation, we call the following errors as the category A; 
detection efficiency, energy resolution, and efficiency in muon charge 
identification. 
%
%For purpose of illuminating the properties of errors of different characters,  
We assume that the dominant part of the category A errors are 
correlated between $\nu$ and $\bar{\nu}$ channels. 
This assumption can be relaxed, if necessary.

\begin{itemize}

\item

$\sigma_{PB}$ ($\nu-\bar{\nu}$ and bin-by-bin correlated error):
Uncertainties in detector volume, 
energy-independent component of the category A  errors.

\item

$\sigma_{Pb}$ ($\nu-\bar{\nu}$  correlated but bin-by-bin uncorrelated error):
Energy-dependent component of the category A  errors.

\item

$\sigma_{pB}$  ($\nu-\bar{\nu}$  uncorrelated but bin-by-bin correlated error):
Uncertainties in beam energy estimate, neutrino flux estimate,  
energy-independent part of reaction cross section error.

\item

$\sigma_{pb}$ (uncorrelated error):
Energy- and channel-dependent fluctuation of the category A  errors, etc., 
which should be very small. 

\end{itemize}

We take, without any solid informations for these errors at this moment, 
$\sigma_{PB} = \sigma_{Pb}=\sigma_{pB}=2$\% and 
$\sigma_{pb}=1$\% in our analysis. 
The expected small values of the errors are partly based on the 
fact that the uncertainties in the muon energy and the luminosity 
are negligibly small for our present purpose 
\cite{Raja-Tollestrup,nufact_eu-us}. 
We also examine the stability of the results by relaxing these errors by a factor of 2. 
We will see that measurement is still dominated by the statistical error 
and a much smaller uncorrelated error, $\sigma_{pb}=0.1$\%, 
would lead to a minor improvement.

\subsection{Treatment of CP phase $\delta$}  
\label{delta}

Despite that $\delta$-dependence is expected to be absent 
at the magic baseline the proper procedure of the analysis has to 
include varying over $\delta$ during the fit. 
The best way to do this is to carry out a combined analysis of data 
taken by both the intermediate and the far detectors.  
Leaving it to a future work, we give in this paper a simplified treatment 
by mimicking data at the intermediate detector by a constraint 
imposed on $\delta$ by adding the term 
\begin{eqnarray}
\Delta \chi^2_{\delta} = 
\left( \frac{ \delta - \delta_{\text{input}} }{ \sigma_{\delta} } \right)^2. 
\label{chi2-delta}
\end{eqnarray}
in $\Delta \chi^2$ in (\ref{chi2}).  
We have chosen 
$\sigma_{\delta} = 0.35$ (20 degree) as a representative value 
based on the estimation of the width of $\chi^2$ parabola at around 
the local minimum as given in \cite{precisionCP}.

We note that ``by definition'' of the magic baseline 
the procedure of varying over $\delta$ should give no significant effect. 
In fact, we have confirmed that the property holds in a wide region of 
$\theta_{13}$ and $\delta$, apart from the one with a subtle feature 
to be addressed in Sec.~\ref{delta-dep}. 
We have also confirmed that using a factor of 2 larger value of 
$\sigma_{\delta}$ does not produce visible changes in the results 
in the entire region.

\section{Analysis results}
\label{results}

\subsection{Analysis with fixed $\delta$} 
\label{fixed-delta}

We first discuss the results with fixed $\delta$. 
It is to have better understanding of the structure of analysis of 
the far detector and the roles of the systematic errors without 
worrying about influence of $\delta$.

%%%%%%%%%%%% FIGURE 2 %%%%%%%%%%%%%%%
\begin{figure}[bhtp]
%\vglue -0.10cm
\begin{center}
%\vglue  -0.5cm
\includegraphics[width=0.8\textwidth]{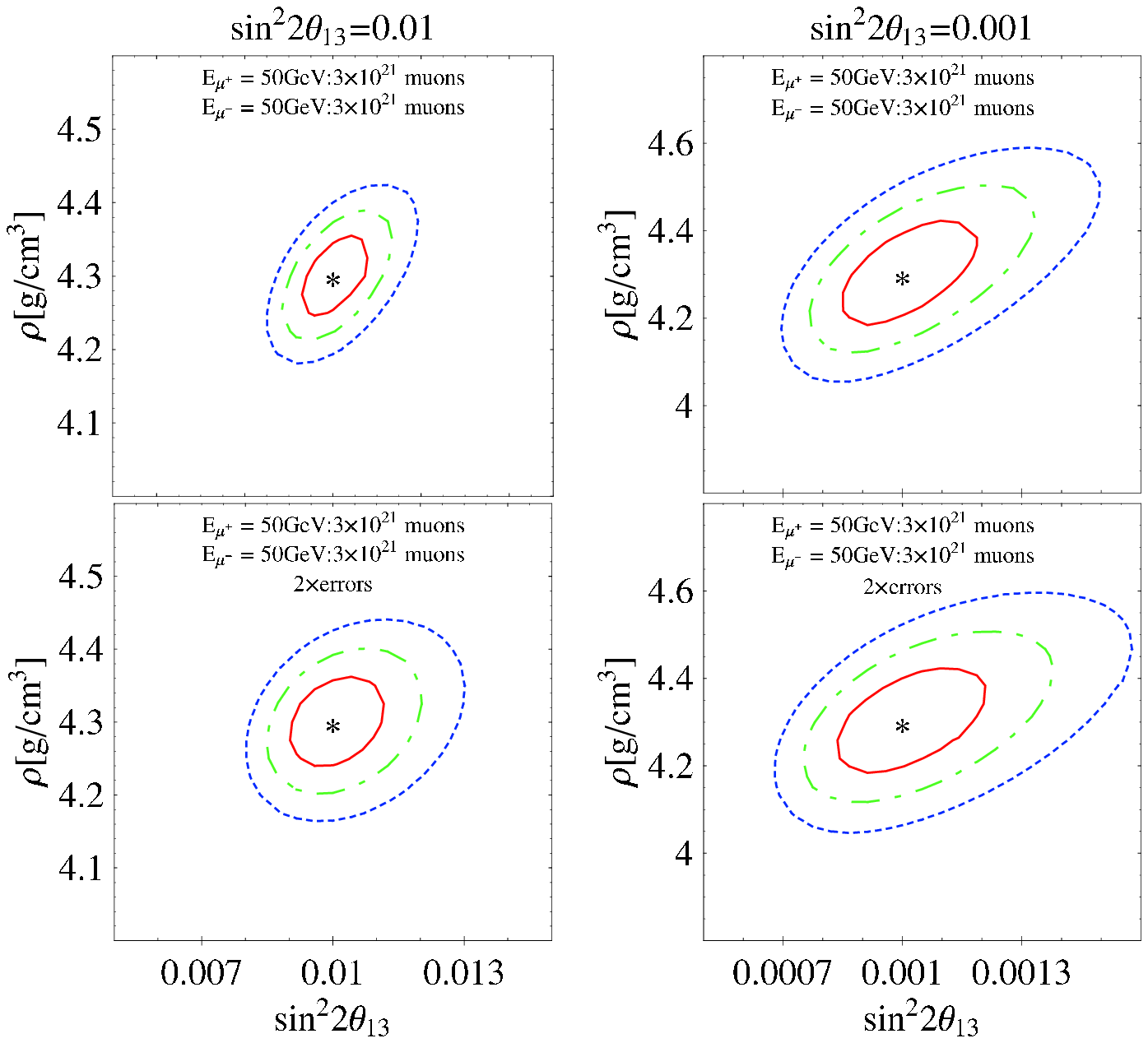}
\end{center}
%\vglue -0.8cm
\caption{Regions in $\sin^2 2 \theta_{13} - \rho$ space allowed by 
operation of neutrino factory with $3 \times 10^{21}$ useful muon 
decays per each polarity watched by 40 kton magnetized iron 
detector at $L=7500$ km. 
The red solid, the green dash-dotted, and the blue dotted lines are for 
contours at 1, 2, and 3 $\sigma$ CL, respectively, defined with 2 DOF. 
The left (right) two panels are for $\sin^2 2 \theta_{13} = 0.01$ (0.001). 
The upper two panels are for the systematic errors 
$\sigma_{PB} = \sigma_{Pb}=\sigma_{pB}=2$\% and 
$\sigma_{pb}=1$\%
as described in the text, and the lower two panels are with a 
factor of 2 larger (apply to all) errors. 
The normal mass hierarchy is assumed and 
CP phase is taken to be $\delta=0$. 
}
\label{2d-plot}
\end{figure}
%%%%%%%%%%%% FIGURE 2 %%%%%%%%%%%%%%%

In Fig.~\ref{2d-plot}, 
presented is the region in $\sin^2 2 \theta_{13} - \rho$ space 
allowed by measurement of $\nu_{\mu}$ and $\bar{\nu}_{\mu}$ 
appearance events with our reference setting of 
neutrino factory. 
The confidence level (CL) contours are defined with 
2 degrees of freedom (DOF) in Fig.~\ref{2d-plot}. 
The input matter density is taken as $\rho = 4.3$ g/cm$^3$. 
The normal mass hierarchy is assumed and 
$\delta$ is taken to be zero. 
The true values of $\sin^2 2 \theta_{13}$ is assumed to be  
0.01 and 0.001 in the left and right two panels, respectively. 
The upper and the lower two panels are for the systematic errors 
given in Sec.~\ref{procedure} and the twice larger values than those, 
respectively. 
Notice that the region is compact not only in $\rho$ direction 
but also in $\sin^2 2 \theta_{13}$ direction despite the fact that 
we have analyzed the data at the far detector only. 
(If the binning of the data were not done the analysis would yield 
a prolonged contour in $\sin^2 2 \theta_{13} - \rho$ space, 
as one expects from (\ref{Pemu}).) 
It is due to a high sensitivity to $\theta_{13}$ at the magic baseline 
which stems from that the probability is free from $\delta$. 
It is impressive to see this behavior because the statistics is quite 
small due to the long baseline, as can be seen in Fig.~\ref{spectrum}.

As indicated in the lower two panels in Fig.~\ref{2d-plot}, 
a factor of 2 enlarged systematic errors gives a relatively 
minor effect on $\delta \rho / \rho$, making the original 
uncertainty 1.3\% (2.8\%) to a modestly larger value 
1.4\% ($^{+2.9}_{-2.8}$\%) both at 1 $\sigma$ CL at 
$\sin^2 2 \theta_{13}=0.01$ (0.001). 
On the other hand, the enlarged systematic errors gives 
a larger effect on $\theta_{13}$, 
worsening the original uncertainty of $\sin^2 2 \theta_{13}$ from 
($^{+8}_{-6}$\%) to ($^{+12}_{-10}$\%) 
[($^{+19}_{-15}$\%) to ($^{+21}_{-16}$\%)] at 1 $\sigma$ CL at 
$\sin^2 2 \theta_{13}=0.01$ [0.001]. 
The fact that a factor of 2 relaxed systematic errors lead to only 
modest increase of $\delta \rho / \rho$ must be emphasized. 
The tendency is more prominent for $\sin^2 2 \theta_{13}=0.001$. 
It is because the detector at the magic baseline is highly sensitive 
to density change and measurement is still dominated by 
the statistical error.

Different roles played by various errors which lead to the enhanced 
uncertainties should be mentioned. 
The factor 2 larger overall normalization error $\sigma_{PB}$ 
primarily affects the uncertainty in $\sin^2 2 \theta_{13}$ 
as expected, and gives little effect to the uncertainty of matter 
density $\delta \rho$. 
The increasing error of $\delta \rho / \rho$ is due to accumulation of 
small and comparable contributions by the other three kind of the 
systematic errors, $\sigma_{Pb}$, $\sigma_{pB}$ and $\sigma_{pb}$. 
The enhanced error of $\sin^2 2 \theta_{13}$, on the other hand, 
is contributed mostly by $\sigma_{PB}$ and $\sigma_{pB}$ 
apart from minor contribution from $\sigma_{Pb}$.

Here are some comments on the effects of including the energy resolution 
and possible values of $\Delta m^2_{31}$ different from the standard one. 
The numbers presented below are for the case of normal mass hierarchy 
but we have checked that the situation is similar in the case of inverted hierarchy. 
The CP phase $\delta$ is taken to be $\delta = 0$. 

\vspace{0.2cm}

\noindent
(1) Because we use only two wide ranged energy bins, it is likely 
that including the energy resolution into our analysis gives only a 
minor effect to the sensitivity to $\delta \rho / \rho$. 
To confirm this point explicitly we have included the energy resolution 
into the analysis by assuming the gaussian form with width 
$\sigma_{E} = 0.15 E$ GeV (as taken in \cite{munich2}). 
At $\sin^2 2 \theta_{13} = 0.01$, $\delta \rho / \rho$ becomes worse from 
0.81\% (without $\sigma_{E}$) to 0.86\%  (with $\sigma_{E}$). 
At $\sin^2 2 \theta_{13} = 0.001$, the corresponding numbers are 
1.8\% (without $\sigma_{E}$) and 1.9\%  (with $\sigma_{E}$). 
Since the effect of $\sigma_{E}$ is so minor we ignore the 
energy resolution in our subsequent analysis. 

\vspace{0.2cm}

\noindent
(2) If the value of $\Delta m^2_{31}$ is larger than 
$\Delta m^2_{31}=2.5 \times 10^{-3}$ eV$^2$ which is assumed 
in our analysis, there is no problem; 
$\delta \rho / \rho$ is comparable or smaller than the 
one obtained above. 
However, If the value of $\Delta m^2_{31}$ is smaller, 
the sensitivity to $\delta \rho / \rho$ becomes less. 
If we assume $\Delta m^2_{31}=1.5 \times 10^{-3}$ eV$^2$, the lower limit 
allowed by the Super-Kamiokande data (the last reference in \cite{SKatm}), 
we have to lower the critical energy $E_{c}$ in defining the energy bins. 
We have used $E_{c} = 8$ GeV and $E_{c} = 10$ GeV for the neutrino 
and the antineutrino channels, respectively. 
The obtained sensitivity to $\delta \rho / \rho$ becomes worse but only 
moderately from the ones quoted above to 1\% and 2\% at 
$\sin^2 2 \theta_{13} = 0.01$ and 0.001, respectively. 
We note, however, that the latest value of $\Delta m^2_{31}$ from the 
MINOS experiment is higher, 
$\Delta m^2_{31}=2.74 \pm^{0.44}_{0.26}$ eV$^2$ (1$\sigma$ CL),
and that the accuracy of determination must be greatly improved by 
the T2K and the NO$\nu$A experiments.

\subsection{Analysis with varied $\delta$} 
\label{varied-delta}

%%%%%%%%%%%% FIGURE 3 %%%%%%%%%%%%%%%
\begin{figure}[bhtp]
%\vglue -0.10cm
\begin{center}
\vglue 0.3cm
\includegraphics[width=0.74\textwidth]{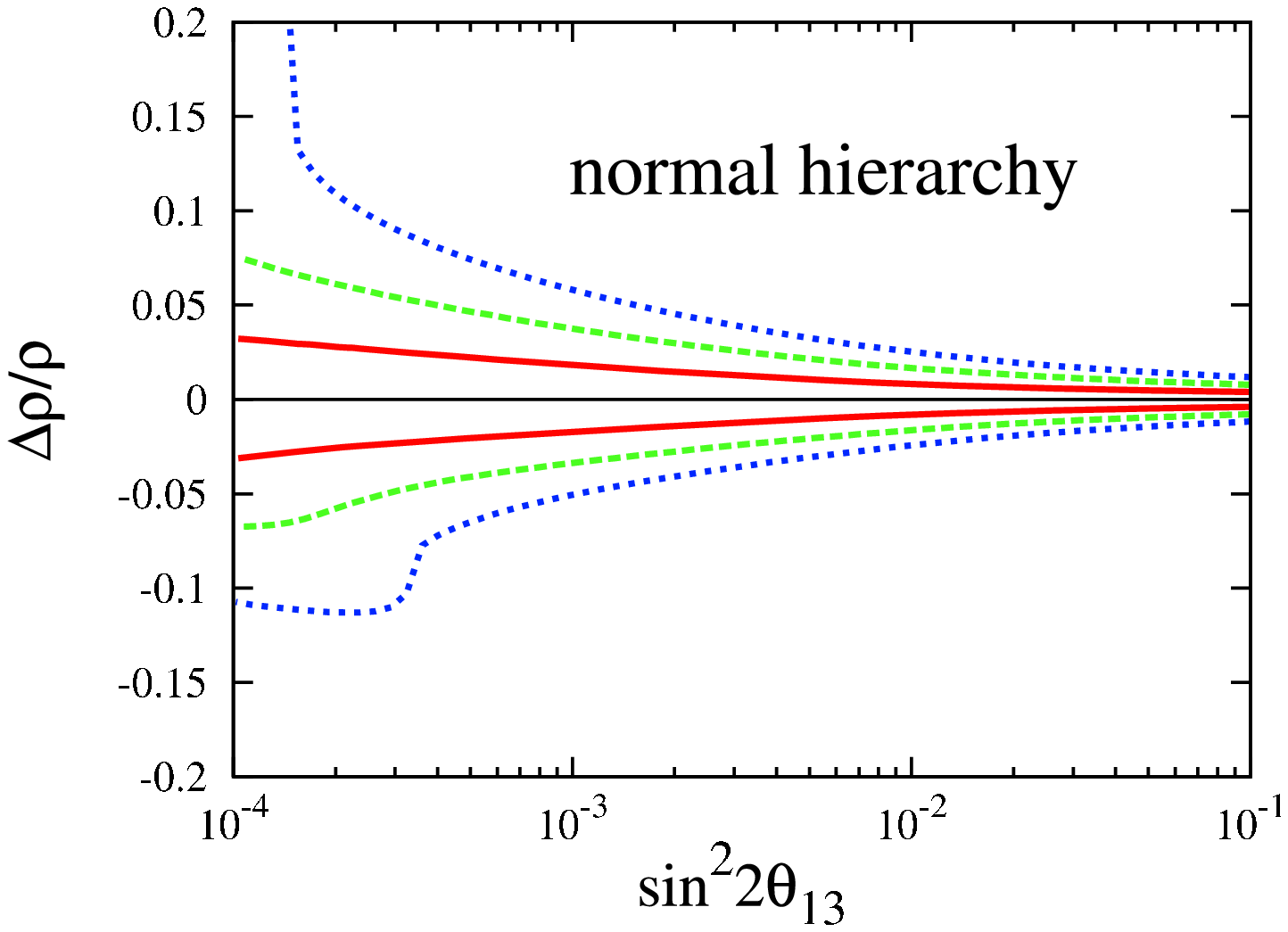}
\vglue  0.5cm
\includegraphics[width=0.74\textwidth]{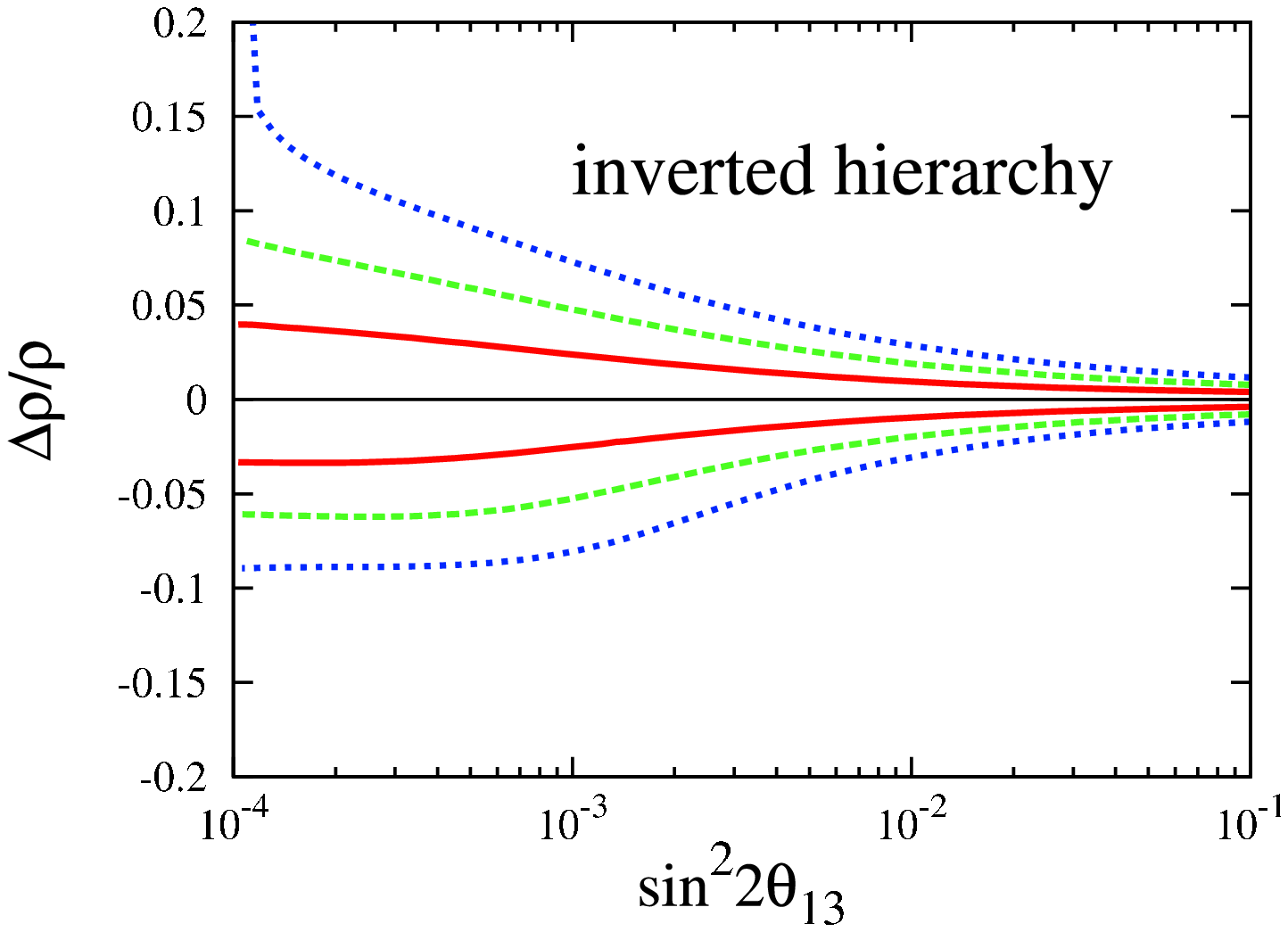}
\end{center}
%\vglue -0.8cm
\caption{The fractional errors in the matter density determination 
$ \delta \rho / \rho$ at 1, 2, and 3 $\sigma$ CL defined with 1 DOF 
by marginalizing $\theta_{13}$ are plotted as a function of 
$\sin^2 2 \theta_{13}$ 
by the red solid, the green dash-dotted, and the blue dotted lines, 
respectively. 
The upper panel is for the normal mass hierarchy with 
$\delta = 0$ and 
the lower panel for the inverted mass hierarchy with 
$\delta = 4\pi/3$. 
}
\label{drho_best}
\end{figure}
%%%%%%%%%%%% FIGURE 3 %%%%%%%%%%%%%%%

We now turn to the analysis with varied $\delta$; 
During the fit we vary $\delta$ with addition of $\Delta \chi^2_{\delta}$
in (\ref{chi2-delta}). 
In Fig.~\ref{drho_best}, we present the fractional error $\delta \rho / \rho$ 
as a function of $\sin^2 2 \theta_{13}$ at 1, 2, and 3 $\sigma$ CL. 
Hereafter, the uncertainty $\delta \rho$ is defined with 1 DOF 
by marginalizing $\sin^2 2 \theta_{13}$. 
The upper and the lower two panels in Fig.~\ref{drho_best} are for 
the normal mass hierarchy with $\delta=0$ and for 
the inverted mass hierarchy with $\delta=4\pi/3$, respectively. 
As we will fully discuss in Sec.~\ref{delta-dep} 
the accuracy is the best at around $\delta=0$ and $\delta=\pi$ 
in the normal and the inverted hierarchies, respectively. 
The case of $\delta=4\pi/3$ is shown in Fig.~\ref{drho_best} to 
indicate that the $\delta$-dependence 
(which exists only in small $\theta_{13}$ region) 
is mild and to avoid to confine ourselves into the 
CP conserving values of $\delta$. 
It should be mentioned here that though we have varied 
$\delta$ for proper analysis procedure it does not produce, 
apart from region of very small $\theta_{13}$, any 
sizable changes in the results, in accord with the natural expectation.

Notice the remarkable accuracy of determination of the 
matter density $\rho$ at the magic baseline represented 
in Fig.~\ref{drho_best}; 
%%%%%%%%%%%% remarkable accuracy %%%%%%%%%%%%%%
The uncertainty is only about 1\% at $\sin^2 2 \theta_{13}=0.01$ 
at 1$\sigma$ CL for both the normal and the inverted mass hierarchies. 
At $\sin^2 2 \theta_{13}=0.001$, the uncertainty increases to 
about 2\% (2.5\%) at 1$\sigma$ CL for the case of the normal (inverted) 
mass hierarchy. 
At $\sin^2 2 \theta_{13}=0.0001$, however, 
$\delta \rho / \rho$ becomes worse to about 3\% (4\%) at 1$\sigma$ CL 
for the case of the normal (inverted) mass hierarchy.

Unfortunately, however, 
we have to mention that a subtlety exists on $\delta$-dependence of 
the sensitivity, which is quite unexpected from the usually advertised 
nature of the magic baseline. It will be fully discussed in the next section.

Before entering into the problem, let us note the followings: 

\vspace{0.2cm}

\noindent
(1) We have explicitly confirmed by plotting $\delta \rho / \rho$ 
as a function of $L$ that distances comparable to the magic baseline 
are the right distances for accurate determination of the earth 
matter density in the energy binning method. 
In fact, the fractional error $| \delta \rho / \rho |$ has broad minima 
in region of distance $L = 7000 - 9000$ km if $\delta$ is fixed, and 
$L = 7500 - 9000$ km if $\delta$ is varied.
See Fig.~5.5 in \cite{uchinami}. 
(The former exercise is to avoid extra complication due to 
$\delta$ dependence which exists in the oscillation probabilities in 
general except for the magic baseline.)

\vspace{0.2cm}

\noindent
(2) We have also investigated the possibility that the accuracy of determination 
of the earth matter density can be improved by doing energy scan 
as well as neutrino energy binning, but without success.

\section{$\delta$ dependence of uncertainty in matter density determination}
\label{delta-dep}

Despite the remarkable accuracy presented in Fig.~\ref{drho_best} 
it is not the end of our work. 
We must address the issue of a rather strong 
$\delta$-dependence in the fractional error $\delta \rho / \rho$ 
at small $\theta_{13}$, which significantly alters the feature of the 
results presented above, as the ``best cases'' with the particular 
values of $\delta$ chosen.

\begin{figure}[bhtp]
\begin{center}
\vglue 0.2cm
\includegraphics[width=0.74\textwidth]{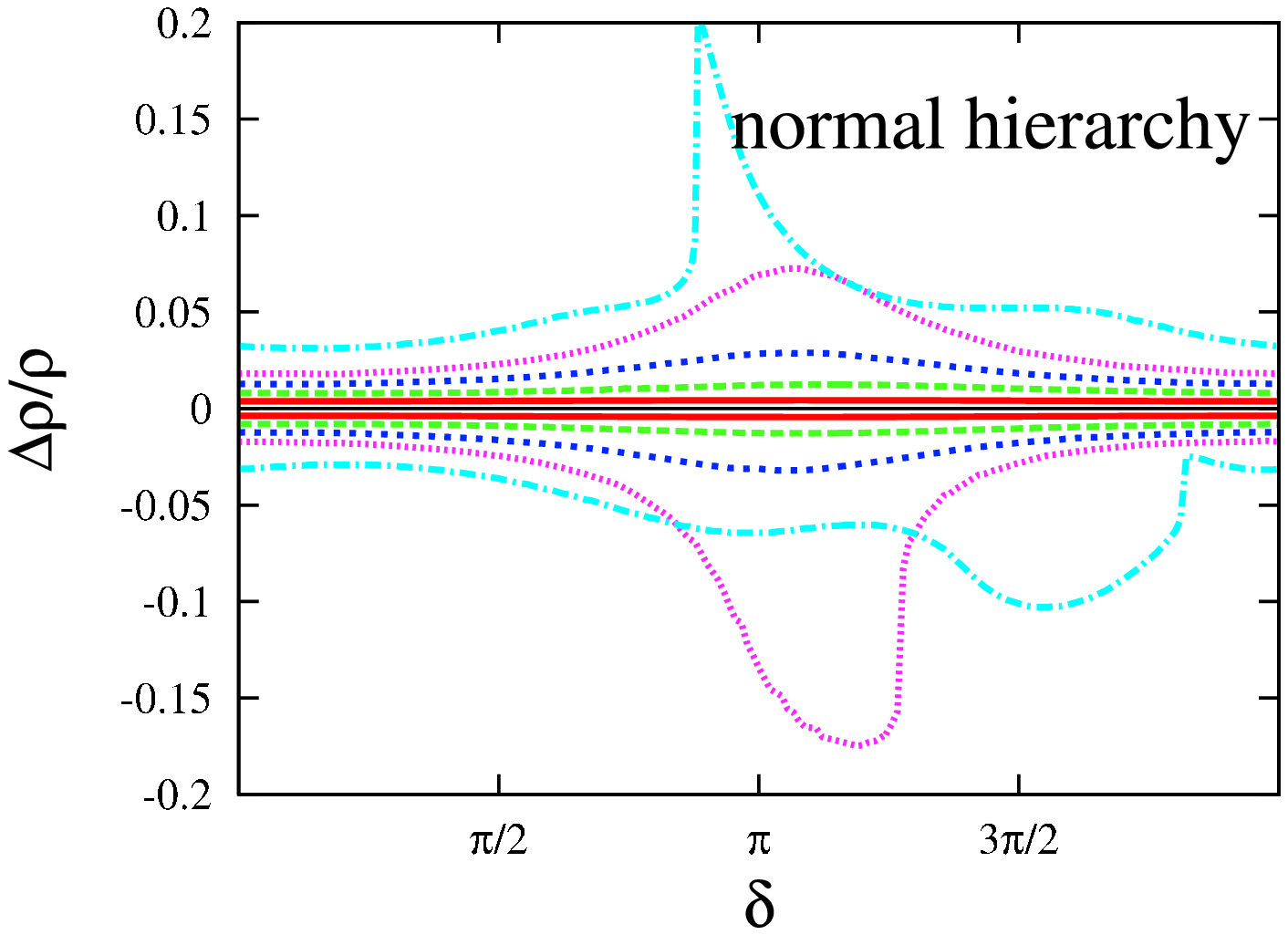}
\vglue  0.4cm
\includegraphics[width=0.74\textwidth]{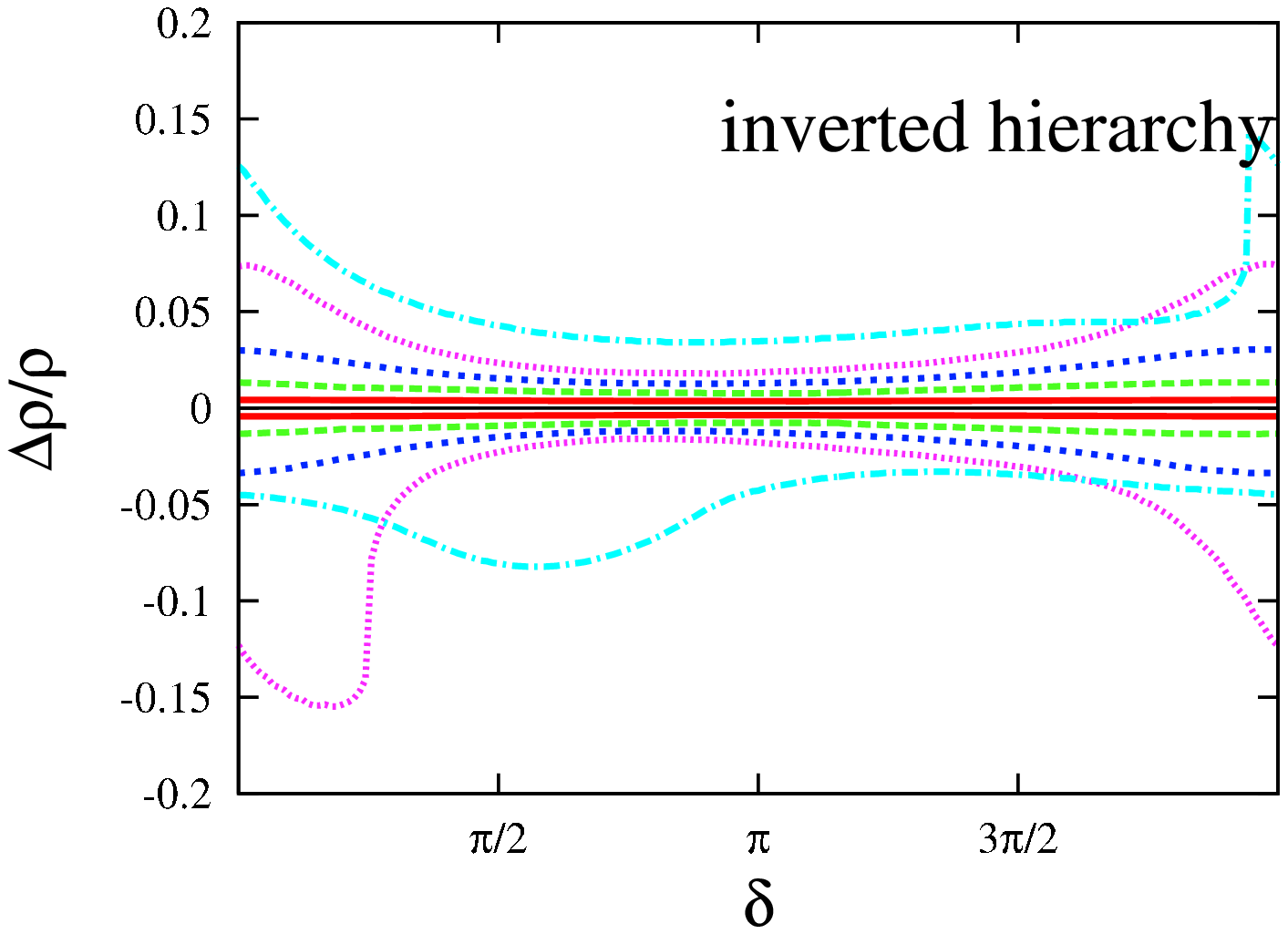}
\end{center}
\vglue -0.1cm
\caption{Presented are the fractional errors $\delta \rho / \rho$ at 
1 $\sigma$ CL with 1 DOF as a function of $\delta$ for five different 
values of $\theta_{13}$, 
$\sin^2 2 \theta_{13} = 0.1$ (red solid line), 
$\sin^2 2 \theta_{13} = 0.01$ (green dashed line), 
$\sin^2 2 \theta_{13} = 0.003$ (blue short-dashed line), 
$\sin^2 2 \theta_{13} = 0.001$ (magenta dotted line), 
$\sin^2 2 \theta_{13} = 0.0001$ (light-blue dash-dotted line). 
The upper and the lower panels are for the normal and the inverted 
mass hierarchies, respectively. 
}
\vglue -0.2cm
\label{drho_delta-dep}
\end{figure}

In Fig.~\ref{drho_delta-dep}, presented is the fractional uncertainty 
$\delta \rho / \rho$ at 1 $\sigma$ CL defined with 1 DOF as a function 
of $\delta$ for four different values of $\theta_{13}$, 
$\sin^2 2 \theta_{13} = 0.1$ (red solid line), 
$\sin^2 2 \theta_{13} = 0.01$ (green dashed line), 
$\sin^2 2 \theta_{13} = 0.003$ (blue short-dashed line), 
$\sin^2 2 \theta_{13} = 0.001$ (magenta dotted line), 
$\sin^2 2 \theta_{13} = 0.0001$  (light-blue dash-dotted line). 
There is a significant differences between the cases of small and large 
$\theta_{13}$ separated by a critical value, 
$\sin^2 2 \theta_{13} \simeq \text{a few} \times 10^{-3}$. 
At larger $\theta_{13}$ than the critical value, 
the sensitivity to matter density is excellent; 
The fractional error $\delta \rho / \rho$ is about 
0.43\%, 1.3\%, and $\lsim$3\% at 1$\sigma$ CL at 
$\sin^2 2\theta_{13}=0.1$, $10^{-2}$, and $3 \times 10^{-3}$, respectively, 
for all values of $\delta$. 
It is also notable that $\delta$-dependence of $\delta \rho / \rho$ is 
very mild in the large $\theta_{13}$ region,  
which confirms the naive expectation due to independence on 
$\delta$ at the magic baseline.

At small $\theta_{13}$, however, there exists a significant 
$\delta$-dependence in $\delta \rho / \rho$. 
Despite that the $\delta \rho / \rho$ remains rather small in most of 
the region of $\delta$, 
about 4\%$-$6\% even at $\sin^2 2\theta_{13}=10^{-4}$, 
there are some ``spike'' structures in 
$\delta \rho / \rho$ at around 
$\delta  \approx \pi$ for the normal hierarchy and 
$\delta  \approx 0$ for the inverted hierarchy around 
which $\delta \rho / \rho$ blows up to 15\%$-$20\%.
They arise when a separated ``island'' that appear in 
1$\sigma$ CL equi-$\chi^2$ contour 
merges with the ``mainland'' 1$\sigma$ CL allowed region. 
In our analysis we do not make a sophisticated treatment to take into 
account separated islands in estimating the errors. 
Therefore, $\delta \rho / \rho$ in regions near the spike is not 
quantitatively reliable. 
On the other hand, the intricate structure of the equi-$\chi^2$ 
contour suggest that it is unstable to inclusion of additional 
informations. 
Namely, a significant improvement of the sensitivity can be expected 
in small $\theta_{13}$ region by a combined analysis of the intermediate 
and the far detectors.

Apart from the small structures there seems to be a relationship 
between the values of $\delta$ with worst sensitivity in the normal 
and the inverted mass hierarchies as 
$\delta(\text{worst})_{normal} \approx \delta(\text{worst})_{inverted} + \pi$. 
The strong $\delta$-dependence of the sensitivity including this feature 
is quite an unexpected subtlety because we usually think of 
the magic baseline as the distance where the CP phase $\delta$ 
plays minor role. 
Therefore, the feature should be understood better 
and it is the reason why we devote the rest of the present section to 
analyzing the subtlety.

%%%%%%%%%%%% FIGURE 3 %%%%%%%%%%%%%%%
%%%%%%%%%%%% FIGURE 4 %%%%%%%%%%%%%%%

The standard form of oscillation probabilities in 
$\nu_{\mu} \rightarrow \nu_e$ and 
$\bar{\nu}_{\mu} \rightarrow \bar{\nu}_e$ 
channels are given by \cite{golden} 
\begin{eqnarray}
P(\nu_e \rightarrow \nu_{\mu}) &=& 
X_{\pm}\sin^2 2\theta_{13} + 
Y_{\pm} \sin 2\theta_{13} \cos {\left( \delta \mp \Delta_{31} \right)} + 
P_{\odot} 
\label{emuP} \\
P(\bar{\nu}_{e} \rightarrow \bar{\nu}_{\mu})  &=& 
\bar{X}_{\pm} \sin^2 2\theta_{13} +  
\bar{Y}_{\pm} \sin 2\theta_{13} \cos {\left(  \delta \pm \Delta_{31} \right)} + 
P_{\odot} \nonumber \\
&=& 
X_{\mp} \sin^2 2\theta_{13} - 
Y_{\mp} \sin 2\theta_{13} \cos {\left(  \delta \pm \Delta_{31} \right)} + 
P_{\odot} 
\label{emuPbar} 
\end{eqnarray}
where the functions $X_{\pm}$ and $Y_{\pm}$ are defined by 
\begin{eqnarray}
X_{\pm} &=&  s^2_{23}
\left[ \frac{\Delta_{31}\sin({aL \mp \Delta_{31}})}{(aL \mp \Delta_{31})} \right]^2, 
\label{X}\\
%\nonumber  \\
%
Y_{\pm} &=& \pm 2\sqrt{X_\pm P_\odot} = 
\pm 2 \sin{2\theta_{12}} c_{23}s_{23}
\left[ \frac{\Delta_{31}\sin({aL \mp \Delta_{31}})}{(aL \mp \Delta_{31})} \right]
\left[ \frac{\Delta_{21}\sin{({aL})}}{aL}\right]
\label{Y}\\
%\nonumber\\
%
P_{\odot} & = & c^2_{23} \sin^2{2\theta_{12}} 
\left[ \frac{\Delta_{21}\sin{({aL})}}{aL}\right]^2
\end{eqnarray}
where $\pm$ indicates the sign of $\Delta m^2_{31}$ and 
$\Delta_{21} \equiv \Delta m^2_{21} L / 4 E $. 
In the last line in (\ref{emuPbar}) we have used the relationship 
\cite{MNprd98}
between the coefficients $X_{\pm}$ and $Y_{\pm}$ in neutrino 
and antineutrino channels, 
$\bar{X}_{\pm} = X_{\mp}$ and 
$\bar{Y}_{\pm} = - Y_{\mp}$.

We perturb the matter density around the one corresponding to 
the magic baseline as $aL = \pi + \epsilon$. 
We expand the neutrino and the antineutrino oscillation 
probabilities by $\epsilon$ and obtain and obtain, to first order in $\epsilon$, 
\begin{eqnarray}
P(\nu_e \rightarrow \nu_{\mu}) &=&  
P(\nu_e \rightarrow \nu_{\mu}; aL = \pi) 
\nonumber \\ 
&-& 
\frac{2 \epsilon}{\pi} 
c_{23} s_{23}  \sin 2\theta_{13} \Delta_{21} \Delta_{31} 
\frac{\sin {\Delta_{31}} }{(\pi \mp \Delta_{31})} 
\nonumber \\ 
&\times& 
\left [
\cos (\delta \mp \Delta_{31}) +  
\frac{\pi \Delta_{31} }{ \Delta_{21} (\pi \mp \Delta_{31} )}
\frac{\sin 2\theta_{13} \tan \theta_{23} }{ \sin 2\theta_{12} } 
\left(  
\frac{\sin {\Delta_{31}} }{ \pi \mp \Delta_{31} } \pm 
\cos  \Delta_{31} 
\right) 
\right],
\label{Pperturb} \\
%\end{eqnarray}
%
%
%\begin{eqnarray}
P(\bar{\nu}_{e} \rightarrow \bar{\nu}_{\mu})  &=& 
P(\bar{\nu}_{e} \rightarrow \bar{\nu}_{\mu}; aL = \pi)
\nonumber \\ 
&-& 
\frac{2 \epsilon}{\pi} 
c_{23} s_{23}  \sin 2\theta_{13} \Delta_{21} \Delta_{31} 
\frac{\sin {\Delta_{31}} }{(\pi \pm \Delta_{31})} 
\nonumber \\ 
&\times& 
\left [
- \cos (\delta \pm \Delta_{31}) +  
\frac{\pi \Delta_{31} }{ \Delta_{21} (\pi \pm \Delta_{31} )}
\frac{\sin 2\theta_{13} \tan \theta_{23} }{ \sin 2\theta_{12} } 
\left(  
\frac{\sin {\Delta_{31}} }{ \pi \pm \Delta_{31} } \mp
\cos  \Delta_{31} 
\right) 
\right]. 
\label{Pbarperturb}
\end{eqnarray}
Suppose that values of the parameters are such that the quantity in 
the square bracket in (\ref{Pperturb}) or (\ref{Pbarperturb}) cancel out. 
Then, the measurement loses the sensitivity to density change, 
leading to an enlarged error in $\delta \rho / \rho$. 
The phenomenon can take place 
only if the two terms have the same order of magnitudes, 
which occurs at small $\theta_{13}$,  
$\sin 2\theta_{13} \approx \Delta_{21} / \Delta_{31} = 
|\Delta m^2_{21} / \Delta m^2_{31}| \simeq 1/30$, which explains why 
the $\delta$-dependent loss of the sensitivity in Fig.~\ref{drho_delta-dep} 
occurs only in region of $\sin^2 2\theta_{13}$ comparable to $\sim 0.001$.

We note that the event statistics is dominated by the neutrino 
channel in the normal mass hierarchy, and by the 
antineutrino channel in the inverted one, one can observe in 
Fig.~\ref{spectrum}. 
Let us focus on these dominant channels. 
Then, we observe a notable regularity; 
The values of $\delta$ which makes the square brackets in 
 (\ref{Pperturb}) (upper sign) and (\ref{Pbarperturb}) (lower sign) 
vanish are related by 
$\cos (\delta _{\nu; \text{normal}} - \Delta_{31} ) + 
\cos (\delta _{\bar{\nu}; \text{inverted}} - \Delta_{31} ) = 0$, or 
$\delta _{\nu; \text{normal}} = \delta_{\bar{\nu}; \text{inverted}} + \pi$. 
It explains the salient feature of Fig.~\ref{drho_delta-dep}, the 
relationship between the $\delta$'s corresponding to the worst sensitivities to 
$\delta \rho / \rho$ in the normal and the inverted mass hierarchies. 
One can also verify that with $\Delta_{31}  \sim 1$ 
the equations of vanishing the order 
$\epsilon$ terms in  (\ref{Pperturb}) and (\ref{Pbarperturb}) 
are roughly consistent with 
$\delta _{\nu; \text{normal}}  \approx \pi$ and 
$\delta _{\nu; \text{inverted}} \approx 0$.

\section{Global fit vs. iterative method}
\label{iterative}

In this paper, we have restricted ourselves to the simplified analysis 
in which information from the intermediate detector is modeled 
by the effective $\chi^2$ as in (\ref{chi2-delta}). 
Clearly one has to engage in the next step a global analysis of 
all the data set taken by the intermediate and the far detectors which 
aims at measuring the earth matter density {\em in situ} as 
a part of the program of determining all the relevant lepton flavor 
mixing parameters.

What is the right way to carry it out?
A straightforward answer to this question may be to marginalize 
(minimize $\chi^2$) with respect to the remaining parameters 
from the one that one want to determine. 
Since the allowed region in $\sin^2 2 \theta_{13} - \rho$ space 
is already rather compact with use of the data of only a far detector 
the method may produce a satisfactory result by just marginalizing 
the $\chi^2$.

Nonetheless, we describe here an alternative iterative method 
for determination of earth matter density in neutrino factory. 
We start from a zeroth-order assumption of the matter density 
taken from estimate (with uncertainties) by geophysical models.
Then, the iterative procedure includes the following two steps:

\begin{itemize}

\item
We assume an $n$-th order assumption of the matter density with 
uncertainties and carry out the analysis to obtain n-th order values of 
$\theta_{13}$ and $\delta$. 

\item
We use $n$-th order values of $\theta_{13}$ and $\delta$ with 
uncertainties to obtain ($n+1$)-th order value of the matter density.  

\end{itemize}

\noindent
Assuming convergence we expect that the method is able to produce 
an improved estimate of the three relevant parameters, 
$\theta_{13}$, $\delta$ and the matter density $\rho$ in a 
self-consistent way.

We note that, irrespective of which analysis method is employed, 
one can expect improvement of the accuracies in 
determination of the matter density. 
Measurement of the intermediate detector should improve 
$\delta \rho / \rho$ directly because of its sensitivity to the earth matter effect, 
in a similar manner as CP phase measurement by the two-detector 
method \cite{MNplb97}.\footnote{
%%%%%%%%%%%%%%%% footnote %%%%%%%%%%%%%%%%
See Fig.~2 in \cite{SNOW06_mina} obtained in the analysis for the 
first reference in \cite{T2KK} for demonstration of this point.
}
We have already mentioned in the previous section the reasons 
why a significant improvement can be expected at small $\theta_{13}$ 
by combined analysis where a intricate structure of 
$\chi^2$ minima entails worsened errors in a limited region of $\delta$.

\section{Concluding remarks}
\label{conclude}

In this paper, we have demonstrated that an {\em in situ} precision 
determination of the earth matter density can be carried out by 
neutrino factory with a detector placed at the magic baseline. 
There are two regions which are separated by a critical value of 
$\theta_{13}$, 
$\sin^2 2\theta_{13} \simeq \text{a few} \times 10^{-3}$. 
In the large $\theta_{13}$ region, the uncertainty $\delta \rho / \rho$ 
has only mild $\delta$-dependence as expected by the desirable 
feature of measurement at the magic baseline. 
The achievable accuracy of matter density determination is excellent; 
The fractional error 
$\delta \rho / \rho$ is about 0.43\%, 1.3\%, and $\lsim$3\%  
at 1$\sigma$ CL at 
$\sin^2 2\theta_{13}=0.1$, $10^{-2}$, and $3 \times 10^{-3}$, respectively. 
It is worth to note that the sensitivity is insensitive against increasing 
the systematic errors by a factor of 2. 
These are the good enough accuracies to solve the problem of 
notorious uncertainty in the neutrino factory data analysis for 
precision determination of $\theta_{13}$ and the CP phase $\delta$ 
in view of Fig.~24 of \cite{munich2}, for example.

In the smaller $\theta_{13}$ region, however, 
we uncovered a subtle $\delta$-dependence in the 
accuracy of determination at small $\theta_{13}$.  
In a certain range of $\delta$ it occurs that the responses of the 
dominant atmospheric term and the solar-atmospheric interference term 
to matter density variation cancel with each other, leading to 
reduced sensitivity (spike structure in Fig.~\ref{drho_delta-dep}) in 
$\delta \rho / \rho$. 
Adding the $\nu_{\mu}$ disappearance channel does not appear to help. 
We were unable to solve this problem within the framework of analysis 
used in this paper. 
We, however, emphasize that $\delta \rho / \rho$ remains small in 
most of the region of $\delta$; At 
$\sin^2 2\theta_{13} = 10^{-4}$ it is 3\%-7\% at 1$\sigma$ CL 
in more than 3/4 of the entire region of $\delta$. 
Furthermore, we expect that simultaneous analysis of the intermediate 
and the far detector data should improve the situation significantly, 
as discussed in Sec.~\ref{delta-dep}. 
If it works it resolves the long debated problem of obscured CP 
violation by the uncertainty of the earth matter density.

It should be stressed, among other things, is that this 
apparatus seems to provide the most accurate way to directly 
measure the matter density inside the earth. 
It will provide a stringent test for geophysical estimation of the 
matter density and/or the MSW theory of neutrino propagation 
in matter.

Finally, we want to mention about other possible ways which 
could provide with us different ways of measuring matter density 
or alternative means of testing the MSW theory.

\begin{itemize}

\item
The transition region of $^8$B neutrinos from the sun in the 
energy region of $3-4$MeV, if measured with precision, 
would allow us to accurately determine the matter density in 
the solar interior. 
Given the fact that we have two alternative ways 
to cross check the results, 
helioseismology and the standard solar model calculation, 
this possibility may give another good way for stringent test of 
the MSW theory of neutrino propagation in matter. 

\item

As one can learn from the expression of the oscillation probability 
(\ref{emuP}) and (\ref{emuPbar}) the relative importance of the 
matter effect to the vacuum effect in neutrino oscillation 
is not controlled completely by the ratio $aL / \Delta_{31}$. 
It may open the possibility of using other variables as tunable parameters, e.g., the baseline $L$. 
It would then be interesting to examine this possibility in the light 
of recent proposal of the Tokai-to-Kamioka-Korea (T2KK) project 
\cite{T2KK,T2KKweb}. Because of the identical two detector setting, 
it would allow us to accurately determine the matter density in the crust 
region below the sea of Japan. 

\item

Finally, we want to note that measuring the matter density and 
confronting it to other method is {\em not} the only way to verify the 
MSW theory. For example, the concept of mass eigenstate in 
matter as well as the regeneration of neutrino flavors are best 
tested by observing the day-night effect in solar neutrino observation.

\end{itemize}

%%%%%%%%%%%%%%%% acknowledgments %%%%%%%%%%%%%
\begin{acknowledgments}
We thank the Theoretical Physics Department of Fermilab for hospitality 
extended to us during the visit in the summer 2005 
when this work was initiated. 
One of the authors (H.M.) is grateful to Maria Gonzalez-Garcia for 
useful discussions in an infant stage of this work. 
He thanks Alexei Smirnov for his illuminating comments and to 
Abdus Salam International Center for Theoretical Physics 
for hospitality. 
The other (S.U.) thanks Olga Mena for discussions about details of 
the analysis in \cite{golden}.  
This work was supported in part by the Grant-in-Aid for Scientific Research, 
No. 16340078, Japan Society for the Promotion of Science.

\end{acknowledgments}

%%%%%%%%%%%%%%%% references %%%%%%%%%%%%%


\begin{thebibliography}{99}


\bibitem {SKatm}
Y.~Fukuda {\it et al.}  [Kamiokande Collaboration],
%``Atmospheric muon-neutrino / electron-neutrino ratio in the multiGeV energy range,''
Phys.\ Lett.\ B {\bf 335}, 237 (1994);
%%CITATION = PHLTA,B335,237;%%
%
Y.~Fukuda {\it et al.}  [Super-Kamiokande Collaboration],
%``Evidence for oscillation of atmospheric neutrinos,''
Phys.\ Rev.\ Lett.\  {\bf 81}, 1562 (1998)
[arXiv:hep-ex/9807003];
%%CITATION = HEP-EX 9807003;%%
%
Y.~Ashie {\it et al.}  [Super-Kamiokande Collaboration],
  %``A measurement of atmospheric neutrino oscillation parameters by
  %Super-Kamiokande I,''
  Phys.\ Rev.\ D {\bf 71}, 112005 (2005)
  [arXiv:hep-ex/0501064].
  %%CITATION = HEP-EX 0501064;%%


\bibitem {solar}
B.~T.~Cleveland {\it et al.},
%``Measurement Of The Solar Electron Neutrino Flux With The Homestake  Chlorine Detector,''
Astrophys.\ J.\  {\bf 496}, 505 (1998);
%%CITATION = ASJOA,496,505;%%
%
J.~N.~Abdurashitov {\it et al.}  [SAGE Collaboration],
%``Measurement of the solar neutrino capture rate with gallium metal,''
Phys.\ Rev.\ C {\bf 60}, 055801 (1999)
[arXiv:astro-ph/9907113];
%%CITATION = ASTRO-PH 9907113;%%
%
W.~Hampel {\it et al.}  [GALLEX Collaboration],
%``GALLEX solar neutrino observations: Results for GALLEX IV,''
Phys.\ Lett.\ B {\bf 447}, 127 (1999);
%%CITATION = PHLTA,B447,127;%%
%
M.~Altmann {\it et al.} [GNO Collaboration],
% "Complete results for five years of GNO solar neutrino observations"
Phys.\ Lett.\ B {\bf 616}, 174 (2005)
[arXiv:hep-ex/0504037];
%
J.~Hosaka {\it et al.}  [Super-Kamiokande Collaboration],
  %``Solar neutrino measurements in Super-Kamiokande-I"
  Phys.\ Rev.\ D {\bf 73}, 112001 (2006)
  [arXiv:hep-ex/0508053];
  %%CITATION = HEP-EX 0508053;%%
%
%M.~B.~Smy {\it et al.}  [Super-Kamiokande Collaboration],
%  %``Precise measurement of the solar neutrino day/night and seasonal  variation
%  %in Super-Kamiokande-I,''
%  Phys.\ Rev.\ D {\bf 69}, 011104 (2004)
%  [arXiv:hep-ex/0309011];
%  %%CITATION = HEP-EX 0309011;%%
%
Q.~R.~Ahmad {\it et al.}  [SNO Collaboration],
%``Measurement of the charged current interactions produced by B-8  solar neutrinos at the Sudbury Neutrino Observatory,''
Phys.\ Rev.\ Lett.\  {\bf 87}, 071301 (2001)
[arXiv:nucl-ex/0106015];
%%CITATION = NUCL-EX 0106015;%%
%Q.~R.~Ahmad {\it et al.}  [SNO Collaboration],
%``Direct evidence for neutrino flavor transformation from neutral-current  interactions in the Sudbury Neutrino Observatory,''
%Phys.\ Rev.\ Lett.\  
{\it ibid.} {\bf 89}, 011301 (2002)
[arXiv:nucl-ex/0204008]; 
%%CITATION = NUCL-EX 0204008;%%
B.~Aharmim {\it et al.}  [SNO Collaboration],
  %``Electron energy spectra, fluxes, and day-night asymmetries of B-8 solar
  %neutrinos from the 391-day salt phase SNO data set,''
  Phys.\ Rev.\ C {\bf 72}, 055502 (2005)
  [arXiv:nucl-ex/0502021].
  %%CITATION = NUCL-EX 0502021;%%


\bibitem{KamLAND}
K.~Eguchi {\it et al.} [KamLAND Collaboration],
Phys.\ Rev.\ Lett.\  {\bf 90}, 021802 (2003) 
[arXiv:hep-ex/0212021]; 
%%CITATION = HEP-EX 0212021;%%


\bibitem{atm_evidence}
  Y.~Ashie {\it et al.}  [Super-Kamiokande Collaboration],
  %``Evidence for an oscillatory signature in atmospheric neutrino
  %oscillation,''
  Phys.\ Rev.\ Lett.\  {\bf 93}, 101801 (2004)
  [arXiv:hep-ex/0404034].
  %%CITATION = HEP-EX 0404034;%%


\bibitem{KL_evidence}
  T.~Araki {\it et al.}  [KamLAND Collaboration],
  %``Measurement of neutrino oscillation with KamLAND: Evidence of spectral
  %distortion,''
  Phys.\ Rev.\ Lett.\  {\bf 94}, 081801 (2005)
  [arXiv:hep-ex/0406035].
  %%CITATION = HEP-EX 0406035;%%


\bibitem{K2K_evidence}
  E.~Aliu {\it et al.}  [K2K Collaboration],
  %``Evidence for muon neutrino oscillation in an accelerator-based
  %experiment,''
  Phys.\ Rev.\ Lett.\  {\bf 94}, 081802 (2005)
  [arXiv:hep-ex/0411038].
  %%CITATION = HEP-EX 0411038;%%


\bibitem{MINOS}
D.~G.~Michael {\it et al.}  [MINOS Collaboration],
  %``Observation of muon neutrino disappearance with the MINOS detectors and
  %the NuMI neutrino beam,''
  Phys.\ Rev.\ Lett.\  {\bf 97}, 191801 (2006)
  [arXiv:hep-ex/0607088].
  %%CITATION = PRLTA,97,191801;%%


\bibitem {MNS}
Z.~Maki, M.~Nakagawa and S.~Sakata,
%``Remarks On The Unified Model Of Elementary Particles,''
Prog.\ Theor.\ Phys.\  {\bf 28}, 870 (1962).
%%CITATION = PTPKA,28,870;%%


\bibitem {superbeam}
The idea of conventional superbeam for measuring CP violation 
may be traced back to: 
%
H. Minakata and H. Nunokawa, Phys. Lett. {\bf B495}, 369 (2000) 
[arXiv:hep-ph/0004114];
%%CITATION = HEP-PH 0004114;%%
J. Sato, Nucl. Instrum. Meth. {\bf A472}, 434 (2001)
[arXiv:hep-ph/0008056];
%%CITATION = HEP-PH 0008056;%%
B. Richter, arXiv:hep-ph/0008222.
%%CITATION = HEP-PH 0008222;%%


\bibitem {nufact}
S.~Geer,
%``Neutrino beams from muon storage rings: Characteristics and physics
%potential,''
Phys.\ Rev.\ D {\bf 57}, 6989 (1998)
[Erratum-ibid.\ D {\bf 59}, 039903 (1999)]
[arXiv:hep-ph/9712290]; 
%%CITATION = HEP-PH 9712290;%%
A.~De Rujula, M.~B.~Gavela and P.~Hernandez,
Nucl. Phys. {\bf B547}, 21 (1999)  
[arXiv:hep-ph/9811390]. 
%%CITATION = HEP-PH 9811390;%%


\bibitem{beta}
P.~Zucchelli,
%``A novel concept for a anti-nu/e / nu/e neutrino factory: The beta beam,''
Phys.\ Lett.\ B {\bf 532}, 166 (2002).
%%CITATION = PHLTA,B532,166;%%


\bibitem {nufact_eu-us}
M.~Apollonio {\it et al.}, arXiv:hep-ph/0210192; 
%%CITATION = HEP-PH 0210192;%%
%
C.~Albright {\it et al.},
  %``Physics at a neutrino factory,''
  arXiv:hep-ex/0008064.
  %%CITATION = HEP-EX 0008064;%%


\bibitem{beta2}
 J.~Bouchez, M.~Lindroos and M.~Mezzetto,
  %``Beta beams: Present design and expected performances,''
  AIP Conf.\ Proc.\  {\bf 721}, 37 (2004)
  [arXiv:hep-ex/0310059].
  %%CITATION = HEP-EX 0310059;%%


\bibitem{international}
A.~Baldini {\it et al.}  [BENE Steering Group],
   ``Beams for European Neutrino Experiments (BENE): Midterm scientific report'' 
\href{http://www.slac.stanford.edu/spires/find/hep/www?irn=6619533}{SPIRES entry}; \\
C.~Albright {\it et al.}  [Neutrino Factory/Muon Collider Collaboration],
  %``The neutrino factory and beta beam experiments and development,''
  arXiv:physics/0411123.
  %%CITATION = PHYS-ICS 0411123;%%


\bibitem{golden}
  A.~Cervera, A.~Donini, M.~B.~Gavela, J.~J.~Gomez Cadenas, P.~Hernandez, O.~Mena and S.~Rigolin,
  %``Golden measurements at a neutrino factory,''
  Nucl.\ Phys.\ B {\bf 579}, 17 (2000)
  [Erratum-ibid.\ B {\bf 593}, 731 (2001)]
  [arXiv:hep-ph/0002108].
  %%CITATION = HEP-PH 0002108;%%
  

\bibitem{wolfenstein}
L.~Wolfenstein,
Phys.\ Rev.\ D {\bf 17}, 2369 (1978).
%%CITATION = PHRVA,D17,2369;%%


\bibitem{MS}
S.~P.~ Mikheyev and A.~Yu.~ Smirnov,
Yad.\ Fiz.\ {\bf 42}, 1441 (1985)
[ Sov.\ J. Nucl.\ Phys.\ {\bf 42}, 913 (1985)];
Nuovo Cim.\ C {\bf 9}, 17 (1986).
%%CITATION = NUCIA,C9,17;%%


\bibitem {AKS}
J.~Arafune, M.~Koike and J.~Sato,
%{\it CP violation and matter effect in long baseline neutrino oscillation  experiments,
Phys.\ Rev. \ D {\bf 56} (1997) 3093
[{\it Erratum-ibid.}\ D {\bf 60} (1997) 119905], 
[arXiv:hep-ph/9703351].
%%CITATION = HEP-PH 9703351;%%


\bibitem {MNprd98}
H.~Minakata and H.~Nunokawa,
%   ``CP violation vs. matter effect in long-baseline neutrino oscillation experiments,''
  Phys.\ Rev.\ D {\bf 57}, 4403 (1998)
  [arXiv:hep-ph/9705208]; 
  %%CITATION = HEP-PH 9705208;%%


\bibitem{Freund}
  M.~Freund, M.~Lindner, S.~T.~Petcov and A.~Romanino,
%   ``Testing matter effects in very long baseline neutrino oscillation experiments,''
  Nucl.\ Phys.\ B {\bf 578}, 27 (2000)
  [arXiv:hep-ph/9912457].
  %%CITATION = HEP-PH 9912457;%%


\bibitem{MNjhep01}
 H.~Minakata and H.~Nunokawa,
  %``Exploring neutrino mixing with low energy superbeams,''
  JHEP {\bf 0110}, 001 (2001)
  [arXiv:hep-ph/0108085].
  %%CITATION = HEP-PH 0108085;%%


\bibitem{Joe}
M.~Koike, T.~Ota and J.~Sato,
%   ``Ambiguities of theoretical parameters and CP/T violation in neutrino factories,''
  Phys.\ Rev.\ D {\bf 65}, 053015 (2002)
  [arXiv:hep-ph/0011387].
  %%CITATION = HEP-PH 0011387;%%


\bibitem{Burguet-C}
  J.~Burguet-Castell, M.~B.~Gavela, J.~J.~Gomez-Cadenas, P.~Hernandez and O.~Mena,
  %``On the measurement of leptonic CP violation,''
  Nucl.\ Phys.\ B {\bf 608}, 301 (2001)
  [arXiv:hep-ph/0103258].
  %%CITATION = HEP-PH 0103258;%%
  

\bibitem{yasuda}
J.~Pinney and O.~Yasuda,
%  ``Correlations of errors in measurements of CP violation at neutrino factories,''
  Phys.\ Rev.\ D {\bf 64}, 093008 (2001)
  [arXiv:hep-ph/0105087].
  %%CITATION = HEP-PH 0105087;%%
  

\bibitem{munich1}
 P.~Huber, M.~Lindner and W.~Winter,
  %``Superbeams versus neutrino factories,''
  Nucl.\ Phys.\ B {\bf 645}, 3 (2002)
  [arXiv:hep-ph/0204352].
  %%CITATION = HEP-PH 0204352;%%


\bibitem{Ohlsson}
  T.~Ohlsson and W.~Winter,
%   ``The role of matter density uncertainties in the analysis of future neutrino factory experiments,''
  Phys.\ Rev.\ D {\bf 68}, 073007 (2003)
  [arXiv:hep-ph/0307178].
  %%CITATION = HEP-PH 0307178;%%
  

\bibitem{munich2}
P.~Huber, M.~Lindner, M.~Rolinec and W.~Winter,
  %``Optimization of a neutrino factory oscillation experiment,''
  arXiv:hep-ph/0606119.
  %%CITATION = HEP-PH 0606119;%%


\bibitem{geller-hara}
  R.~J.~Geller and T.~Hara,
  %``Geophysical aspects of very long baseline neutrino experiments,''
  Nucl.\ Instrum.\ Meth.\ A {\bf 503}, 187 (2001)
  [arXiv:hep-ph/0111342].
  %%CITATION = HEP-PH 0111342;%%


\bibitem{Cervera_nufact06}
A.~Cervera, 
Talk given at Eighth International Workshop on the Neutrino Factories, 
Superbeams, and Beta Beams, University of California, Irvine, USA, 
August 24-30, 2006.


\bibitem{Panasyuk} 
For Reference Earth Model, see e.g., the website; \\
http://cfauvcs5.harvard.edu/lana/rem/index.htm


\bibitem{bari_focus}
G.~Fogli and E.~Lisi,
  %``Evidence for the MSW effect,''
  New J.\ Phys.\  {\bf 6}, 139 (2004).
  %%CITATION = NJOPF,6,139;%%


\bibitem{tomography_beam}
W.~Winter,
%   ``Direct test of the MSW effect by the solar appearance term in beam experiments,''
  Phys.\ Lett.\ B {\bf 613}, 67 (2005)
  [arXiv:hep-ph/0411309].
  %%CITATION = HEP-PH 0411309;%%


\bibitem{tomography_core}
W.~Winter,
%   ``Probing the absolute density of the earth's core using a vertical neutrino beam,''
  Phys.\ Rev.\ D {\bf 72}, 037302 (2005)
  [arXiv:hep-ph/0502097].
  %%CITATION = HEP-PH 0502097;%%
  

\bibitem{tomography_SN}
  M.~Lindner, T.~Ohlsson, R.~Tomas and W.~Winter,
  %``Tomography of the earth's core using supernova neutrinos,''
  Astropart.\ Phys.\  {\bf 19}, 755 (2003)
  [arXiv:hep-ph/0207238].
  %%CITATION = HEP-PH 0207238;%%
  

\bibitem{tomography_rev}
For an extensive bibliography, see 
 W.~Winter,
%   ``Neutrino tomography: Learning about the earth's interior using the propagation of neutrinos,''
  arXiv:physics/0602049.
  %%CITATION = PHYS-ICS 0602049;%%


\bibitem{MEMPHYS}
J.~E.~Campagne, M.~Maltoni, M.~Mezzetto and T.~Schwetz,
  %``Physics potential of the CERN-MEMPHYS neutrino oscillation project,''
  arXiv:hep-ph/0603172.
  %%CITATION = HEP-PH 0603172;%%


\bibitem {T2K}
Y.~Itow {\it et al.}, arXiv:hep-ex/0106019.\\
%%CITATION = HEP-EX 0106019;%%
For an updated version, see:
http://neutrino.kek.jp/jhfnu/loi/loi.v2.030528.pdf


\bibitem{nufact01_mina}
  H.~Minakata and H.~Nunokawa,
  %``CP trajectory diagram: A tool for pictorial representation of CP and
  %matter effects in neutrino oscillations,''
  Nucl.\ Instrum.\ Meth.\ A {\bf 503}, 218 (2001)
  [arXiv:hep-ph/0111130].
  %%CITATION = HEP-PH 0111130;%%


\bibitem{T2KK}
  M.~Ishitsuka, T.~Kajita, H.~Minakata and H.~Nunokawa,
  %``Resolving neutrino mass hierarchy and CP degeneracy by two identical
  %detectors with different baselines,''
  Phys.\ Rev.\ D {\bf 72}, 033003 (2005)
  [arXiv:hep-ph/0504026]; 
  %%CITATION = HEP-PH 0504026;%%
T.~Kajita, H.~Minakata, S.~Nakayama and H.~Nunokawa,
  %``Resolving eight-fold neutrino parameter degeneracy by two identical
  %detectors with different baselines,''
  Phys.\ Rev.\  D {\bf 75}, 013006 (2007)
  [arXiv:hep-ph/0609286].
  %%CITATION = PHRVA,D75,013006;%%


\bibitem{magicBL}
P.~Huber and W.~Winter,
  %``Neutrino factories and the 'magic' baseline,''
  Phys.\ Rev.\ D {\bf 68}, 037301 (2003)
  [arXiv:hep-ph/0301257].
  %%CITATION = HEP-PH 0301257;%%
  

\bibitem{smirnov_magic}
A.~Y.~Smirnov,
  %``Neutrino oscillations: What is magic about the 'magic' baseline?,''
  arXiv:hep-ph/0610198. 
  %%CITATION = HEP-PH 0610198;%%
For earlier discussions, see e.g., A.~Y.~Smirnov,
  %``The MSW effect and matter effects in neutrino oscillations,''
  Phys.\ Scripta {\bf T121}, 57 (2005)
  [arXiv:hep-ph/0412391].
  %%CITATION = HEP-PH 0412391;%%


\bibitem{uchinami}
S.~Uchinami, 
Master of Science Thesis, Tokyo Metropolitan University, February 2007 (in Japanese), 
http://musashi.phys.metro-u.ac.jp/thesis.html


\bibitem{BMW}
V.~Barger, D.~Marfatia and K.~Whisnant,
%   ``Breaking eight-fold degeneracies in neutrino CP violation, mixing, and mass hierarchy,''
  Phys.\ Rev.\ D {\bf 65}, 073023 (2002)
  [arXiv:hep-ph/0112119].
  %%CITATION = HEP-PH 0112119;%%


\bibitem{KTY}
K.~Kimura, A.~Takamura and H.~Yokomakura,
  %``Exact formula of probability and CP violation for neutrino oscillations  in
  %matter,''
  Phys.\ Lett.\  B {\bf 537}, 86 (2002)
  [arXiv:hep-ph/0203099].
  %%CITATION = PHLTA,B537,86;%%


\bibitem{Gandhi}
  R.~Gandhi, P.~Ghoshal, S.~Goswami, P.~Mehta and S.~Uma Sankar,
  %``Large matter effects in nu/mu --> nu/tau oscillations,''
  Phys.\ Rev.\ Lett.\  {\bf 94}, 051801 (2005)
  [arXiv:hep-ph/0408361]; 
  %%CITATION = HEP-PH 0408361;%%
 %R.~Gandhi, P.~Ghoshal, S.~Goswami, P.~Mehta and S.~Uma Sankar,
  %``Earth matter effects at very long baselines and the neutrino mass
  %hierarchy,''
  Phys.\ Rev.\ D {\bf 73}, 053001 (2006)
  [arXiv:hep-ph/0411252].
  %%CITATION = HEP-PH 0411252;%%


\bibitem{Blondel}
A.~Blondel, A.~Cervera-Villanueva, A.~Donini, P.~Huber, M.~Mezzetto and P.~Strolin,
  ``Future neutrino oscillation facilities,''
  arXiv:hep-ph/0606111.
  %%CITATION = HEP-PH 0606111;%%


\bibitem{Long_nufact06}
K.~Long, Talk given at Eighth International Workshop on the Neutrino Factories, Superbeams, and Beta Beams, University of California, Irvine, USA, 
August 24-30, 2006.


\bibitem{Ellis_nufact06}
M.~Ellis,
Talk given at Eighth International Workshop on the Neutrino Factories, 
Superbeams, and Beta Beams, University of California, Irvine, USA, 
August 24-30, 2006.


\bibitem {reactorCP}
H.~Minakata and H.~Sugiyama, 
Phys.\ Lett.\ {\bf B580}, 216 (2004) 
[arXiv:hep-ph/0309323]. 
%%CITATION = HEP-PH 0309323;%%


\bibitem{Raja-Tollestrup}
  R.~Raja and A.~Tollestrup,
  %``Calibrating the energy of a 50-GeV x 50-GeV muon collider using spin
  %precession,''
  Phys.\ Rev.\ D {\bf 58}, 013005 (1998) 
  [arXiv:hep-ex/9801004].
  %%CITATION = HEP-EX 9801004;%%


\bibitem{precisionCP}
  P.~Huber, M.~Lindner and W.~Winter,
  %``From parameter space constraints to the precision determination of the
  %leptonic Dirac CP phase,''
  JHEP {\bf 0505}, 020 (2005)
  [arXiv:hep-ph/0412199].
  %%CITATION = HEP-PH 0412199;%%


\bibitem {MNplb97}
H.~Minakata and H.~Nunokawa,
%``How to measure CP violation in neutrino oscillation experiments?,''
Phys.\ Lett.\ B {\bf 413}, 369 (1997)
[arXiv:hep-ph/9706281].
%%CITATION = HEP-PH 9706281;%%


\bibitem{SNOW06_mina}
 H.~Minakata,
  %``Resolving degeneracy in neutrino oscillation measurement of mixing
  %parameters,''
  Phys.\ Scripta {\bf 127}, 73 (2006).
  %%CITATION = PHSTB,127,73;%%


\bibitem{T2KKweb}
For broader aspects of ``T2KK'' visit the web pages: \\
http://t2kk.snu.ac.kr/ (2nd International Workshop 
on a Far Detector in Korea for the J-PARC Neutrino Beam, July 2006); \\
http://newton.kias.re.kr/~hepph/J2K/ 
(1st International Workshop, November 2005). 



\end{thebibliography}
\end{document}